\documentclass[aps,prb,showpacs,superscriptaddress,twocolumn]{revtex4}
\usepackage[dvips]{graphicx}
\usepackage{amssymb,amsmath,graphicx}
\usepackage{bm}
\usepackage{ulem}
\usepackage{color}

\newcommand{\be}{\begin{equation}}
\newcommand{\ee}{\end{equation}}
\newcommand{\ba}{\begin{eqnarray}}
\newcommand{\ea}{\end{eqnarray}}
\newcommand{\bi}{\begin{itemize}}
\newcommand{\ei}{\end{itemize}}

\begin{document}

\title{Finite difference method for transport of two-dimensional massless Dirac fermions \\ in a ribbon geometry}

\author{Alexis R. Hern\'andez}
\affiliation{Instituto de F\'{\i}sica, Universidade Federal do Rio de Janeiro, 
21941-972 Rio de Janeiro, Brazil}
\affiliation{Departamento de F\'{\i}sica, Pontif\'{\i}cia Universidade Cat\'olica  do Rio de Janeiro, 
22452-970 Rio de Janeiro, Brazil}
\author{Caio H. Lewenkopf}
\affiliation{Instituto de F\'{\i}sica, Universidade Federal Fluminense, 24210-346 Niter\'oi, Brazil}

\date{\today}

\begin{abstract}
We present a numerical method to compute the Landauer conductance of non-interacting two-dimensional massless Dirac fermions in disordered systems. The method allows for the introduction of boundary conditions at the ribbon edges and to account for an external magnetic field. By construction, the proposed discretization scheme avoids the fermion doubling problem. The method does not rely on an atomistic basis and is particularly useful to deal with long-range disorder, whose correlation length largely exceeds the underlying material crystal lattice spacing. As an application, we study the case of monolayer graphene sheets with zigzag edges subjected long range disorder, which can be modeled by a single-cone Dirac equation. 
\end{abstract}


\maketitle
\section{Introduction}

The growing research activity on the physical properties of graphene \cite{CastroNeto09} as well as on topological insulators \cite{Hasan2010,Qi2011} has increased the interest in methods to solve the two dimensional massless Dirac equation for various settings. In pristine graphene, two Dirac cones placed at the corners of the Brillouin zone, frequently called valleys, provide a good description of  the electronic single-particle low-energy spectrum. For topological insulators, two-dimensional massless Dirac excitations appear as edge states between a non trivial tridimensional (3D) topological insulator and a trivial vacuum.

In distinction to analytical studies, that describe the single-particle properties of graphene by an effective two-dimensional Dirac Hamiltonian, almost all numerical analysis employ an atomistic basis from the onset (for reviews, see for instance Refs.~\onlinecite{Mucciolo2010,DasSarma2011}). Numerical studies of the electronic transport in graphene sheets and ribbons typically use the tight-binding approximation to express the relevant Green's functions in terms of atomic site positions. \cite{Rycerz2007,Lewenkopf2008,RGF4ribbons} This basis choice has the advantage of simplicity, but its computational cost increases very fast with system size. For the cases where short range disorder is important and an accurate description of the system at small length scales is necessary, the use of an atomic basis is not only the simplest, but also  the most natural choice. The picture is very different in  systems characterized by long range disorder and/or subjected to a smooth background electrostatic potential. In these cases, intervalley scattering is absent \cite{CastroNeto09,Mucciolo2010} and the system description in terms of a single Dirac cone is very accurate. Here, a numerical method that starts with the Dirac Hamiltonian is certainly computationally more efficient and very likely to be more insightful. The development of such method is the main goal of this paper. 

Topological insulators \cite{Hasan2010,Qi2011} have rekindled the interest on the research related to concepts of topological order, often used to characterize the singular behavior of the integer and fractional quantum Hall effects. Topological insulators are mainly characterized by the presence of chiral boundary states. It was recently discovered a series of next generation 3D topological insulators. 
These materials exhibit a very simple band structure with a single Dirac cone. The non trivial topological nature of these materials makes their boundary states more robust against disorder than the corresponding ones in graphene. In lattice quantum field theory the introduction of boundary conditions has been guided by the index theorem \cite{Atiyah1968} and by the phenomenological presence of Adler-Bell-Jackiw (chiral) anomalies\cite{AdlerBellJackiw} observed in high energy experiments. In this context, it has been shown that nonlocal boundary conditions are necessary to reproduce the observed anomaly. For graphene and topological insulators the boundary conditions are local and well understood. \cite{BreyFertig06,Bardarson10}

In this study we combine two finite difference methods, \cite{susskind77,stacey82} developed in the context of lattice gauge theory, to numerically compute the scattering matrix of two-dimensional massless Dirac particles. Our method avoids the fermion doubling problem and allows to introduce suitable boundary conditions at the system edges. To illustrate its utility we consider boundary conditions mimicking the zigzag edges of a graphene sheet.\cite{BreyFertig06} Inspired on the procedure proposed by Tworzydlo and collaborators, \cite{Tworzydlo08} we develop a method to compute  transport properties of massless Dirac particles in systems with different edge boundary conditions. In addition, the method we put forward also allows to account for the presence of an external magnetic field.  

We apply the method to compute the average conductivity at the charge neutrality point of large disordered graphene systems. We consider diagonal disorder and compute the average conductivity by taking disorder ensemble averages. We scale the conductivity with the system size to numerically obtain the corresponding universal curve, a matter of an intense controversy in the recent literature.\cite{Mucciolo2010,Ostrovsky2007,Nomura2007,Bardarson2007} We also study the conductance fluctuations at the charge neutrality point, providing numerical evidence of a non universal behavior. 

The presentation is organized as follows. In Sec. \ref{sec:general} we review the discretization strategies used to solve the Dirac Hamiltonian. In Sec.~\ref{sec:discretization} we construct a discretization procedure to calculate the Landauer conductance of two-dimensional massless Dirac particles confined to a ribbon with boundary conditions that mimic zigzag edges in graphene. Ballistic transport is discussed in Sec.~\ref{sec:ballistic}, where we apply the method to reproduce the standard results found in the literature. In Sec.~\ref{sec:diffusive} we apply the transfer matrix method to study diffusive transport in long-range disordered graphene stripes. Next, we show how to include a magnetic field in the method and compute the conductance at the (anomalous) quantum Hall regime in Sec.~\ref{sec:magnetic}. Finally, we present our conclusions and outlook in Sec.~\ref{sec:conclusion}.

\section{Massless Dirac equation in a lattice}
\label{sec:general}

Let us consider the two-dimensional massless Dirac Hamiltonian
\be
\label{eq:H_Dirac}
H=-i\hbar v(\sigma_x\partial_x+\sigma_y\partial_y)+U({\bm r}),
\ee
where $v$ is the velocity of the massless Dirac fermions, $\sigma_x$ and $\sigma_y$ are Pauli matrices, and $U({\bm r})$ is the  potential energy landscape. The eigenvalue problem reads
\begin{equation}
H \Psi = E \Psi \;,
\end{equation}
where $E$ is the fermion energy and $\Psi$ is a two-component (spinor) wave function of the form
\be
\label{eq:spinor}
\Psi=\left(\begin{array}{c}\psi \\ \widetilde{\psi}\end{array}\right)\;.
\ee

For graphene, the two components of the spinor $\Psi$ correspond to the electron wave function amplitudes at each of the two intercalated triangular lattices, denoted by $A$ and $B$, that form the honeycomb lattice.  Equation~\eqref{eq:H_Dirac} is the effective low energy graphene Hamiltonian for electrons in the vicinity of the Brillouin zone $K$-point, called $K$-valley. By replacing $\sigma_y \rightarrow - \sigma_y$ one obtains the corresponding equation for the $K^\prime$-valley. 
\cite{CastroNeto09}

In the case of topological insulators, the spinor components of $\Psi$ are interpreted as the electron $z$-axis spin projection.

Our aim is to find a suitable lattice representation for the $H$ operator in Eq.\ \eqref{eq:H_Dirac}. The square lattice, that is very efficient to discretize a two-dimensional Schr\"odinger Hamiltonian, would be the most natural candidate. However, it is well established in the lattice gauge theory literature \cite{susskind77,stacey82,Bender83} that this choice leads to wrong results. 

Let us understand this statement and motivate the discretization scheme we propose by discussing the typical problems that arise in the discretization of the Dirac equation.  For the sake of simplicity, let us consider the simplest possible setting, namely, a one-dimensional stationary massless fermion field. The Dirac equation reads 
\be
\label{eq:Dirac1D}
-i \hbar v\sigma_x \partial_x \Psi (x) = E \Psi(x) \;.
\ee
Despite the simplicity of  Eq.~\eqref{eq:Dirac1D}, its solution by discretization requires some care. 
The naive evaluation of the derivative as $\partial_x \Psi (x)= [\Psi (x+\Delta)-\Psi (x)]/\Delta$ is unsuited, since it produces a non-Hermitian Hamiltonian. The reason is that the left and the right hand sides of Eq.~\eqref{eq:Dirac1D} are evaluated at different points, namely $x$ and $x+\Delta/2$.
The evaluation of $\partial_x \Psi (x)$ in Eq.~\eqref{eq:Dirac1D} by taking $\partial_x \Psi (x)=[\Psi (x+\Delta)-\Psi (x-\Delta)]/2\Delta$ also fails. In this case, one finds an Hermitian Hamiltonian, but with a wrong spectrum: The number of fermions in the first Brillouin zone is doubled. This is the so-called fermion doubling problem. It arises because the lattice size $\Delta$ is half of the spacing between the points used to estimate $\partial_x \Psi (x)$.\cite{stacey82}

To correctly solve Eq.~\eqref{eq:Dirac1D} by discretization, it is necessary to evaluate $\partial_x \Psi (x)$ and $\Psi (x)$ at the same mesh position and to use a single effective lattice spacing. Below, we review two different  strategies that satisfy these requirements.

{\it Stacey discretization}:\cite{stacey82} The fermionic fields are determined at the lattice points $x_m = m\Delta$ and the Dirac equation is evaluated at the lattice midpoints, namely, $m\Delta + \Delta/2$.

In other words, since the derivative is $\partial_x \Psi (x+\Delta/2)= [\Psi (x+\Delta)-\Psi (x)]/\Delta$ and both sides of Eq.\ \eqref{eq:Dirac1D} must be evaluated at the same position, the fermionic fields are approximated by the average between adjacent sites, namely, $\Psi(x+\Delta/2)=[\Psi (x+\Delta)+\Psi (x)]/2$.

The resulting system of coupled finite difference equations reads
\be
- \frac{i\hbar v }{\Delta}\left(\begin{array}{c}\widetilde{\psi}_{m+1} - \widetilde{\psi}_{m} \\ \psi_{m+1}-\psi_{m}\end{array}\right)\;=
\frac{E}{2} \left(\begin{array}{c}\psi_{m+1} + \psi_{m} \\ \widetilde{\psi}_{m+1}+\widetilde{\psi}_{m}\end{array}\right)\;,
\ee
where $\psi_m=\psi(m\Delta)$ and $\widetilde{\psi}_m=\widetilde{\psi}(m\Delta)$.

{\it Susskind discretization}:\cite{susskind77} This scheme is based on a slightly more involved construction than the one just presented. The fermionic fields, ${\psi}$ and ${\widetilde \psi}$, are defined in two intercalated  lattices.  The lattices have a lattice parameter $\Delta$ and are offset by a distance $\Delta/2$. 

Let us consider the spinor component $\psi$ to be defined at $x_m=m \Delta$ and the component ${\widetilde \psi}$ defined at $\widetilde{x}_m=m\Delta - \Delta/2$. Given that $\sigma_x$ is non diagonal, one can evaluate the upper component of Eq.~\eqref{eq:Dirac1D} at $x$ and the lower component at $x-\Delta/2$. 

The Dirac equation, Eq.~\eqref{eq:Dirac1D}, is approximated by the coupled finite difference equations
\be
-\frac{i\hbar v }{\Delta}\left(\begin{array}{c}\widetilde{\psi}_{m+1} - \widetilde{\psi}_m \\ \psi_m-\psi_{m-1}\end{array}\right) = E \left(\begin{array}{c}\psi_m \\ \widetilde{\psi}_m\end{array}\right)\,,
\ee
where $\psi_m=\psi (m\Delta)$ and ${\widetilde \psi}_m={\widetilde \psi}(m\Delta-\Delta/2 )$. Notice that the definitions of $\psi_m$ and $\widetilde{\psi}_m$ for the Susskind discretization differ from the ones used in  the Stacey discretization scheme.   

In summary, both reviewed discretization procedures fulfill the desired goals: They render Hermitian Hamiltonians with the correct spectrum at the continuum limit, eliminating the fermion doubling problem.

\section{Transfer matrix discretization}
\label{sec:discretization}

In this section we describe how to combine the transfer matrix method and the discretization schemes presented above to  compute the Landauer conductance for graphene ribbons with zigzag edges. We use the Susskind discretization along the ribbon transversal direction, because it allows for the implementation of zigzag-like boundary conditions at the ribbon edges.  Along the longitudinal direction, we use the Stacey scheme, since it can be nicely combined with the transfer matrix method. 

We consider a geometry where source and drain are placed at two opposite sides of a rectangular sheet. The graphene sheet has a length $L$ in the direction connecting source and drain, that is longitudinal to the electronic propagation, and a transversal width $W$.  

We take the $x$-axis along the longitudinal direction, and write Eq.\ \eqref{eq:H_Dirac} as
\be
\label{eq:Dirac2}
\partial_x \Psi = (-i\sigma_z\partial_y-i\sigma_x V/\Delta)\Psi
\ee
where $V=(U-E)\Delta/\hbar v$. The Landauer conductance is obtained by solving Eq.~\eqref{eq:Dirac2} in a lattice using the transfer matrix method. For notational convenience, the lattice size $\Delta$ is accounted for in the definition of $V$, a dimensionless quantity. 

The graphene zigzag-like boundary conditions are implemented by making $\psi=0$ at $y=0$ and $\widetilde{\psi}=0$ at $y=W$, corresponding to opposite edges of the ribbon.\cite{BreyFertig06}  In Appendix \ref{app:armchair} we show how to implement the finite difference method to address armchair-like boundary conditions.

\begin{figure}[!h]
\vskip0.2cm
\includegraphics[width=0.7\columnwidth]{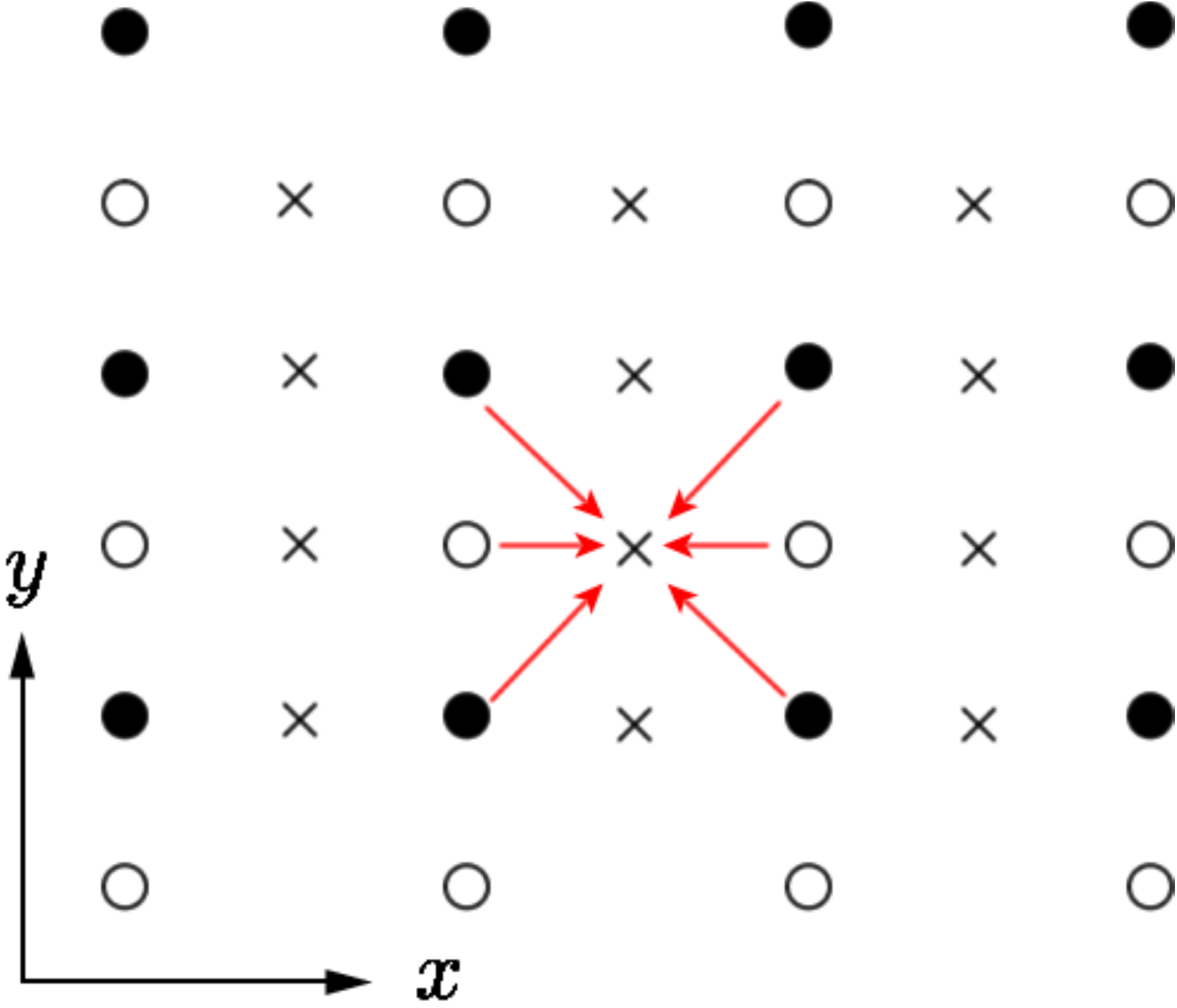}
\caption[]{Representation of the discretization lattice. The $\psi$ spinor is evaluated at the positions corresponding to the filled circles, while $\widetilde \psi$ is evaluated at the open ones. The crosses represent the intercalated lattice where the Dirac equation is evaluated. 
\label{fig:lattice}}
\end{figure}

The discretization of Eq.~\eqref{eq:Dirac2} requires the use of, at least, 3 intercalated lattices. The discretization we put forward in this study is shown in Fig.~\ref{fig:lattice}. We recall that, for graphene, the spinor components $\psi$ and $\widetilde{\psi}$ stand for (continuum) wave function amplitudes at the sublattices $A$ and $B$, respectively. Accordingly, the discrete representation of the spinor component $\psi$ is given by the value of $\psi$ at sites of the lattice $A$  (indicated by filled circles). Likewise, $\widetilde \psi$ is discretized by taking its value at the sites of the lattice $B$ (open circles). The Dirac equation is evaluated at the auxiliary intercalated lattice, represented by crosses in Fig.~\ref{fig:lattice}. \cite{TI}

The finite difference representation of the partial derivatives appearing in  Eq.~\eqref{eq:Dirac2} is given by
\begin{align}
\label{eq:discretization1}
\partial_x \psi =& \frac{1}{2\Delta}(\psi_{m+1,n}-\psi_{m,n})+(\psi_{m+1,n+1}-\psi_{m,n+1}) \\
\partial_y \psi =& \frac{1}{2\Delta}(\psi_{m+1,n+1}-\psi_{m+1,n})+(\psi_{m,n+1}-\psi_{m,n}).\nonumber 
\end{align}
Here, $\partial_x \psi$ and $\partial_y \psi$ are evaluated at the site $s$ of the auxiliary intercalated lattice, located at the center of a square cell whose corners correspond to the lattice $A$ sites where $\psi$ is evaluated.  The diagonal arrows in Fig.~\ref{fig:lattice} illustrate the input required to compute the partial derivatives of $\psi$ at a given site $s$ (indicated by a cross) of the auxiliary lattice. 

Analogously, the expressions for the lattice $B$ read
\begin{align}
\label{eq:discretization2}
\partial_x \widetilde{\psi} =& \frac{1}{2\Delta}(\widetilde{\psi}_{m+1,n}-\tilde{\psi}_{m,n})+(\widetilde{\psi}_{m+1,n-1}-\widetilde{\psi}_{m,n-1})  \\
\partial_y \widetilde{\psi} =& \frac{1}{2\Delta}(\widetilde{\psi}_{m+1,n}-\tilde{\psi}_{m+1,n-1})+(\widetilde{\psi}_{m,n}-\widetilde{\psi}_{m,n-1}) \;.\nonumber
\end{align}
Finally, the lattice approximation for $V\Psi$ in Eq.~\eqref{eq:Dirac2} is 
\begin{equation}
V\psi = V_{m,n}\frac{\psi_{m,n} + \psi_{m+1,n}}{2},
\end{equation} 
which is evaluated, as above, at the site $s$ of the auxiliary lattice. More precisely, $V_{m,n}$ gives the value of $V$ at the midpoint between the sites $(m,n)$ and $(m+1,n)$ of the lattice $A$. The evaluation of $V\psi$ is represented by the horizontal arrows in Fig.~\ref{fig:lattice}. 

For the spinor component $\widetilde{\psi}$ 
\begin{equation}
\label{eq:discreteV2}
V\widetilde{\psi} =\widetilde{V}_{m,n}\frac{\widetilde{\psi}_{m,n} + \widetilde{\psi}_{m+1,n}}{2},
\end{equation} 
gives the value of $V\widetilde{\psi}$ at the auxiliary lattice site $s$ at the midpoint between the sites $(m,n)$ and $(m+1,n)$ of the lattice $B$. 

Throughout this paper, we adopt the convention that the $m$ index is associated with the $x$-axis, while $n$ labels the mesh points $y$-coordinate. For notational convenience, we define the column vectors $ {\bm \psi}_m \equiv (\psi_{m,1}, \cdots, \psi_{m, N})^T$ and $\widetilde{\bm \psi}_m \equiv (\widetilde{\psi}_{m,1}, \cdots, \widetilde{\psi}_{m, N})^T$. We also introduce the column vector
\begin{equation}
{\bm \Psi}_m = \left(\begin{array}{c} {\bm \psi}_m \\ \widetilde{\bm \psi}_m \end{array}\right). 
\end{equation}
that represents the spinor $\Psi$ of Eq.~\eqref{eq:spinor} evaluated at both $A$ and $B$ sublattices for a given $m$, corresponding to a $x=m\Delta$. In Fig.~\ref{fig:lattice}, single columns represent  ${\bm \Psi}_m$ spinors of different $m$'s.

\begin{widetext}

Plugging \eqref{eq:discretization1} to \eqref{eq:discreteV2} in Eq.~\eqref{eq:Dirac2} we obtain
\begin{align}
\label{eq:DisDirac}
({\mathbb I}+{\mathbb N}^{+})({\bm \psi}_{m+1}-{\bm \psi}_{m})= 
&i({\mathbb I}-{\mathbb N}^{+})({\bm \psi}_{m+1}+{\bm \psi}_{m})-i\widetilde{\mathbb V}_{m}( \widetilde{{\bm\psi}}_{m+1}+ \widetilde{{\bm\psi}}_{m})  \\
({\mathbb I}+{\mathbb N}^{-})({\bm {\widetilde \psi}}_{m+1}-{\bm {\widetilde \psi}}_{m})=
&i({\mathbb I}-{\mathbb N}^{-})
( {\widetilde {\bm\psi}}_{m+1}+ {\widetilde {\bm\psi}}_{m})-i{\mathbb V}_{m}({\bm \psi}_{m+1}+{\bm \psi}_{m}) \nonumber
\end{align}
where ${\mathbb I}$ is the $N\times N$ identity matrix,
${\mathbb N}^+$ and ${\mathbb N}^-$ are super- and subdiagonal matrices, namely, 
$({\mathbb N}^+)_{nn^{\prime}}=\delta_{n,n^{\prime}-1}$, $({\mathbb N}^-)_{nn^{\prime}}=\delta_{n,n^{\prime}+1}$, while ${\mathbb V}_m$ and $\widetilde{\mathbb V}_m$ have matrix elements given by $({\mathbb V}_m)_{nn^{\prime}}=V_{mn}\delta_{nn^\prime}$ and $(\widetilde{{\mathbb V}}_m)_{nn^{\prime}}=\widetilde{V}_{mn}\delta_{nn^\prime}$.


%
%
By writing Eq.~\ref{eq:DisDirac} in terms of ${\bm \Psi}_m$ and ${\bm \Psi}_{m+1}$, 
\be
\left( \begin{array}{cc}({\mathbb I} + {\mathbb N}^+)-i({\mathbb I} - {\mathbb N}^+) & i\widetilde{\mathbb V}^m \\
i{\mathbb V}^m & ({\mathbb I} + {\mathbb N}^-)-i({\mathbb I} - {\mathbb N}^-)\end{array}\right)
{\bm \Psi}_{m+1}=
\left( \begin{array}{cc}({\mathbb I} + {\mathbb N}^+)+i({\mathbb I} - {\mathbb N}^+) & -i\widetilde{\mathbb V}^m \\
-i{\mathbb V}^m & ({\mathbb I} + {\mathbb N}^-)+i({\mathbb I} - {\mathbb N}^-)\end{array}\right)
{\bm \Psi}_{m}.
\ee
\end{widetext}
allows us to identify the transfer matrix ${\mathbb M}$ defined by
\be
{\bm \Psi}_{m+1}={\mathbb M}_{m} {\bm \Psi}_{m},
\ee
namely,
\be
\label{eq:M}
{\mathbb M_m}=\frac{{\mathbb I}-i{\mathbb X}_m}{{\mathbb I}+i{\mathbb X}_m},
\ee
where
\be
\label{eq:X}
{\mathbb X}_m={\mathbb J}^{-1} \left( \begin{array}{cc}-({\mathbb I} - {\mathbb N}^+) & \widetilde{\mathbb V}^m \\
{\mathbb V}^m & -({\mathbb I} - {\mathbb N}^-)\end{array}\right)
\ee
and
\be
\label{eq:J}
{\mathbb J}=\left( \begin{array}{cc}{\mathbb I} + {\mathbb N}^+ & 0 \\
0 & {\mathbb I} + {\mathbb N}^-\end{array}\right).
\ee
We define the current operator in the $x$-direction as
\be
\label{eq:Jx}
{\mathbb J}_x=\frac{1}{2}v\left( \begin{array}{cc}0 & {\mathbb J}_- \\
{\mathbb J}_+ & 0\end{array}\right),
\ee
with ${\mathbb J}_+={\mathbb I} + {\mathbb N}^+$ and ${\mathbb J}_-={\mathbb I} + {\mathbb N}^-$. In Appendix \ref{ap:chargcons} we show that this choice of ${\mathbb J}_x$ guarantees charge conservation. 
As a consequence of the zigzag boundary conditions, our discretization method does not preserve symplectic symmetry. This is in contrast with the method put forward in Ref.~\onlinecite{Tworzydlo08}, that deals with periodic boundary conditions along the $y$-direction and treats infinitely wide systems.

\subsection{Landauer conductance}
\label{sec:Landauer}

The transfer matrix of the whole graphene sheet can, in principle, be computed using ${\mathbb M} = \prod_{m=1}^M {\mathbb M}_{m}$. However, it is well known that this scheme is numerically unstable: It produces both exponentially growing and exponentially decaying eigenvalues. Limited numerical accuracy prevents one from retaining both sets of eigenvalues.\cite{Tworzydlo08} The numerical instability problem is solved by working with compositions of unitary scattering matrices, \cite{Tworzydlo08,TamuraAndo1991} at the expense of introducing additional diagonalizations that slow down the computation. We adopt the scheme proposed in Ref.~\onlinecite{Tworzydlo08} with a small modification, which we briefly describe in what follows.

In order to translate the transfer matrix ${\mathbb M}$ into a unitary scattering matrix
\be
{\mathbb S}=\left( \begin{array}{cc} r & t^{\prime} \\
t & r^{\prime} \end{array}\right)
\ee
one needs to solve the equation
\be
\Phi^{\rm right}_a= {\mathbb M} \Phi^{\rm left}_a
\ee
with
\be
\Phi^{\rm right}_a=\sum_l t_{ab} \phi^+_b, \;\;\;  \;\;\; \Phi^{\rm left}_a=\phi^+_a + \sum_l r_{ab} \phi^-_b\,.
\ee
for each asymptotic propagating channel $a$ at the source (left) or drain (right) contacts. We consider the situation where all channels are open. This is similar to the setting where the contacts have a high doping and a large density of states  to realistically model standard metallic contacts, \cite{Tworzydlo06} minimizing the effect of a contact resistance.


In the large wave vector limit, the states $\phi^{\pm}_a$ ($a=1,2,...,N$)  are given by the $2N$ eigenstates of the current operator ${\mathbb J}_x$, normalized such that each mode carries the same current. 
That is
\be
\phi^{\pm}_a=j^{-1/2}_a\left( \begin{array}{c} \psi_{a} \\
\pm {\bar{\psi}}_{a} \end{array}\right)
\ee
where $\psi_a$ and $j_a$ are given by
\ba
{\mathbb J}_+ \psi_a= j_a {\bar{\psi}}_a \nonumber \\
{\mathbb J}_- {\bar{\psi}}_a= j_a \psi_a
\ea
with $({\bar{\psi}}_a)_n=(\psi_a)_{N+1-n}$.  

To obtain a closed-form expression for ${\mathbb S}$ in terms of ${\mathbb M}$, we write ${\mathbb M}$ in the basis of channels $\phi^{\pm}_k$ through the similarity transformation
\be
\widetilde{{\mathbb M}}_m={\mathbb R} \, {\mathbb M}_m \, {\mathbb R}^{-1}
\ee
where
\be
({\mathbb R}^{-1})_{2N\times 2N}=(\phi^+_1, \cdots ,\phi^+_N,\phi^-_1,\cdots,\phi^-_N).
\ee
This ${\mathbb R}$  is not the same as the one used in Ref.~\onlinecite{Tworzydlo08}: We use a different transversal discretization scheme and ${\mathbb R}$ is defined to facilitate the projection of single-contributions to the transmission from different channels. The later characteristic is very useful for the method implementation, as justified in the next section.
%
%

The spinor degrees of freedom of ${\widetilde{{\mathbb M}}}$ are separated into four $N\times N$ blocks,
\be
\widetilde{{\mathbb M}}=\left( \begin{array}{cc} \widetilde{{\mathbb M}}^{++} & \widetilde{{\mathbb M}}^{+-} \\
\widetilde{{\mathbb M}}^{-+} & \widetilde{{\mathbb M}}^{--} \end{array}\right)
\ee
and follow the same strategy employed in Sec. III.b of Ref. \onlinecite{Tworzydlo08} to write the $S$ matrix in terms of the $\widetilde{{\mathbb M}}$ sub-blocks.
Finally, the conductance is evaluated as usual by the Landauer formula
\be
G= \frac{g_s g_v e^2}{h} T
\ee
where $g_s$ and $g_v$ stand for the spin and valley degeneracies and $T$ is the transmission, namely,  $T={\rm tr}[t^{\dagger}t]$.

\subsection{Spurious mode and its elimination}
\label{sec:spurious}

Let us discuss the properties of the transfer matrix for the simple case of a pristine graphene ribbon with zigzag edges. Due to translation invariance, all transversal slices $m$ are equivalent, and the transfer matrix ${\mathbb M}_{\rm PUC}={\mathbb M}_{m}$, corresponding to the  primitive unit cell, is sufficient to describe the transport.

\begin{figure}[!h]
\begin{center}
\includegraphics[width=0.85\columnwidth]{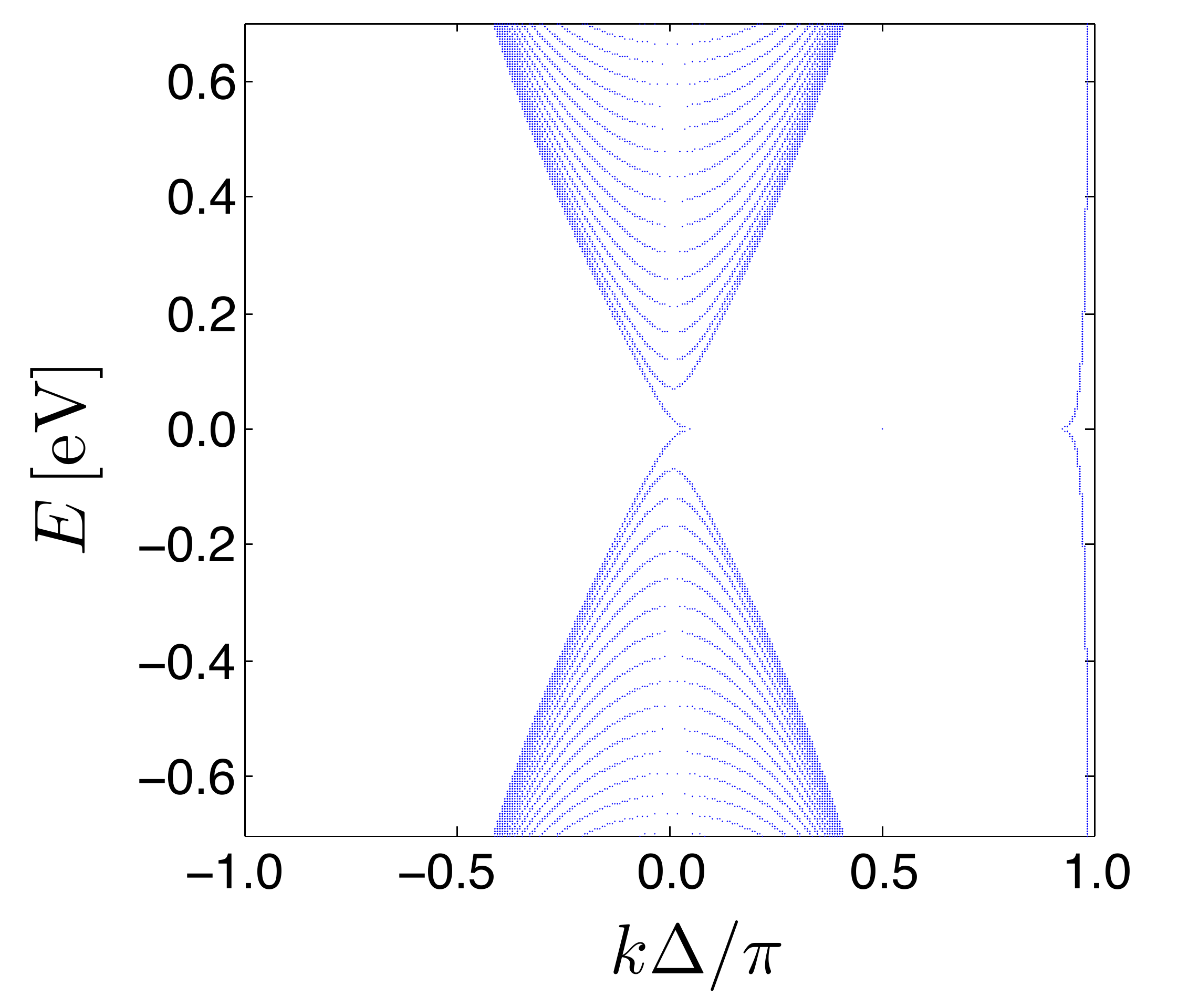}
\end{center}
\vskip-0.5cm
\caption[]{Band structure of Dirac fermions in a infinite ribbon of width $W = 100 \Delta$ with zigzag-like boundary conditions obtained using the method described in the text. The energy scale is set by fixing $\Delta=10 a_0$, where $a_0=1.42$~\AA~is the carbon-carbon bond length in graphene. 
\label{fig:DFbandstructure}}
\end{figure}

For a given energy $E$, the transfer matrix ${\mathbb M}_{\rm PUC}$ has $N_{\rm ch}$  eigenvalues of the form $e^{ik_a \Delta} $ with $k_a$ real. Hence, $N_{\rm ch}$ gives the number of propagating channels and $k_a$ is associated with the wave number along the propagation direction of the corresponding  channel. This allows us to obtain the dispersion relation for a graphene ribbon. For a given ribbon width $W$, we vary the energy $E$ in small steps and calculate the eigenvalues and eigenchannels of  ${\mathbb M}_{\rm PUC}$. The result for the discretization of the $K$-valley is shown in Fig.~\ref{fig:DFbandstructure}.

The band structure shown in Fig.~\ref{fig:DFbandstructure} coincides with the results obtained by the continuum limit treatment of the zigzag boundary conditions of Ref.~\onlinecite{BreyFertig06} around the $K$ valley. As expected, the low energy states are also in good agreement with those of the nearest-neighbor tight-binding model. The only discrepancy is the appearance of an ``extra" mode around $k\Delta/\pi \approx 1$. To understand its nature, it is convenient to analyze the corresponding eigenstate. This is done next.  

\begin{figure}[!h]
\begin{center}
\begin{tabular}{c}
\includegraphics[width=1.0\columnwidth]{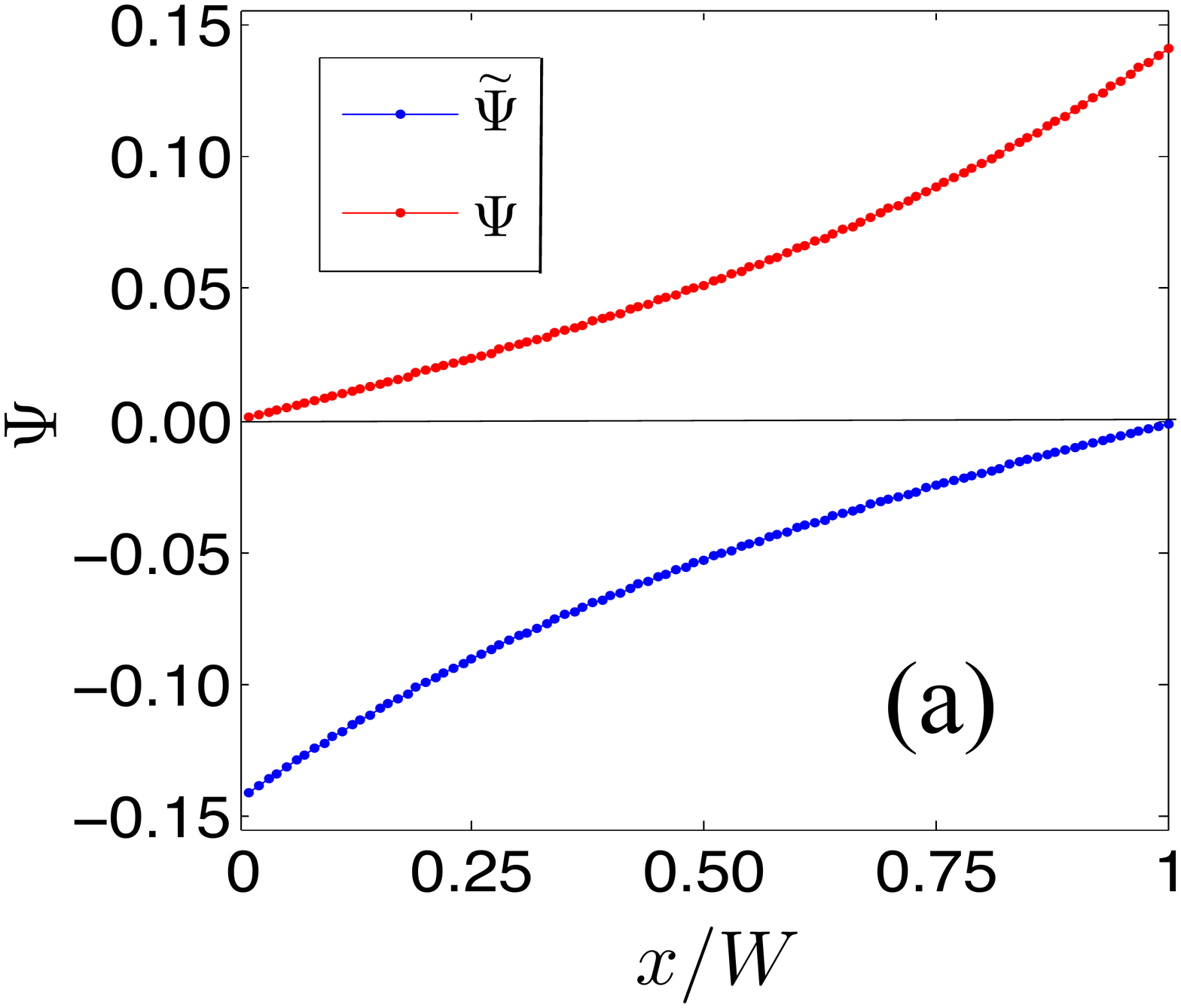} \\
\includegraphics[width=0.95\columnwidth]{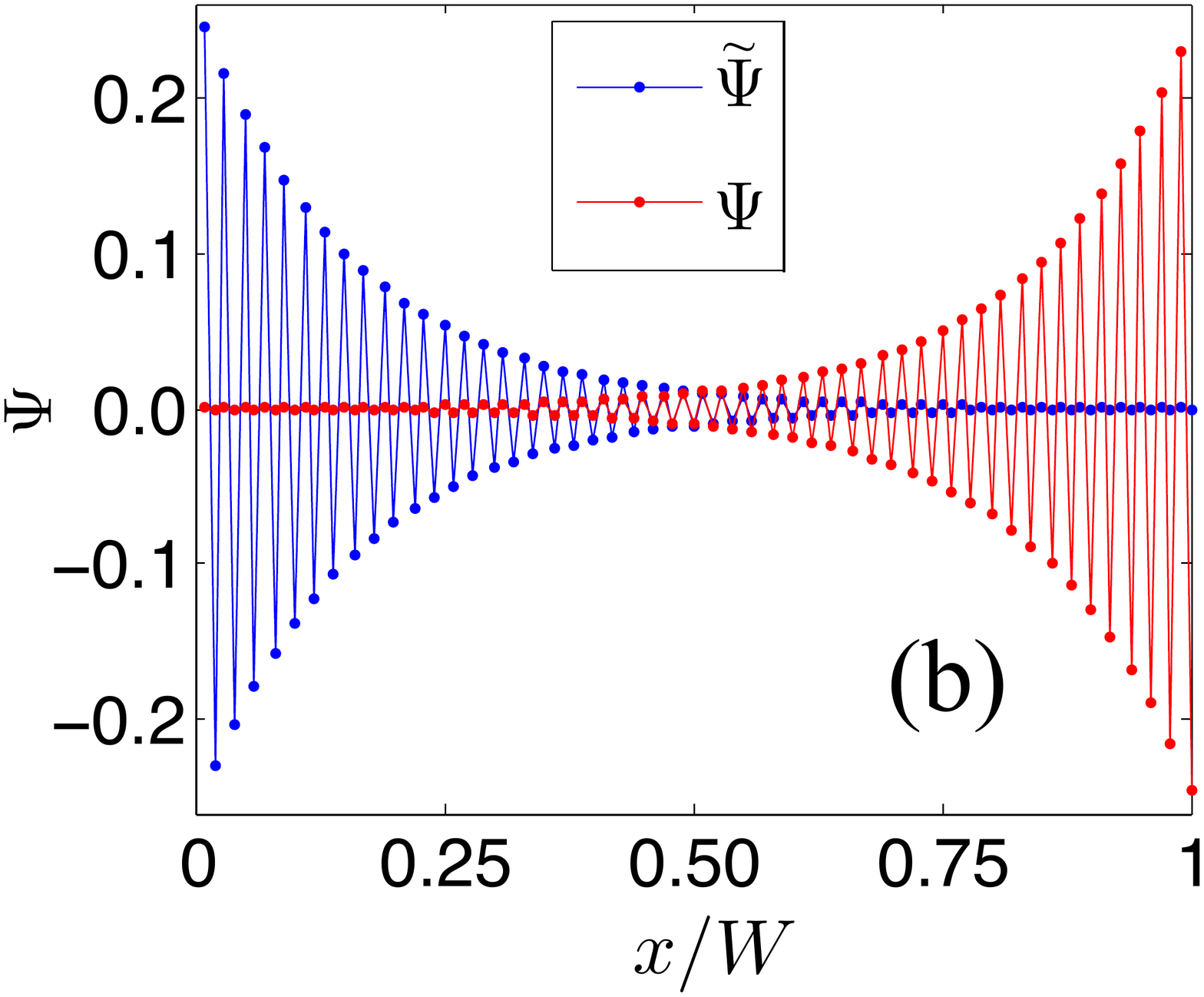}
\end{tabular}
\end{center}
\caption{(Color online) Spinor $\Psi$ projection amplitudes, $\psi$ and $\widetilde{\psi}$, corresponding to the zero-energy  dispersionless modes of ${\mathbb M}_m$: (a) Physical mode and (b) spurious solution.  
\label{fig:modes}}
\end{figure}

The transfer matrix ${\mathbb M}_{\rm PUC}$ has two zero-energy eigenmodes, whose typical transversal wave function amplitudes are shown in Fig.~\ref{fig:modes}.  The ``smooth" mode corresponding to  Fig.~\ref{fig:modes}a is similar to the analytical solution presented in Ref.\ \onlinecite{BreyFertig06}. In addition, we find a spurious mode that oscillates on the scale of the lattice size discretization, shown in Fig.~\ref{fig:modes}b. 

In finite difference calculations, states showing oscillations at the scale of lattice size are expected to be badly described by the discretized equation. Usually, they correspond to the states of higher energy in the spectrum. The general rule is that the first half or so of the computed states are well described by the discretization. Hence, provided one is interested in the low energy part of the spectrum,  the inaccurate high frequency states do not cause any problem. These observations do not apply here. For example, in the Tworzydlo and collaborators \cite{Tworzydlo08}, that employ a Stacey-like discretization to handle the fermion doubling problem, the effective lattice size is doubled to evaluate the  derivatives of ${\bm \Psi}$. As a consequence, the high energy states at the top of the spectrum are poorly described and spurious states arise. To eliminate these spurious states, very long filters at each side of his graphene stripe are introduced. 

In our case, the Susskind discretization preserves the lattice size, however the peculiar nature of the edge state, characteristic of ribbons with zigzag edges, pulls down the energy of the maximally transversal oscillating state. The combined oscillations in both sublattices render a dispersionless mode in the finite difference representation of the momentum operator. 

To eliminate (up to a very good accuracy) the spurious state illustrated in Fig. \ref{fig:modes}b,  we simply eliminate the corresponding channel in definition of ${\mathbb R}$:
\be
({\mathbb R}^{-1})_{2N\times 2N-2}=(\phi^+_1, \cdots ,\phi^+_{N-1},\phi^-_1,\cdots,\phi^-_{N-1}).
\ee
The inversion of the ${\mathbb R}^{-1}$ is taken by transposing the matrix and by inverting the normalization of the eigenvectors $\phi^{\pm}_{a}$. 


\section{Ballistic transport}
\label{sec:ballistic}

In this Section we show that our method reproduces known results for ballistic transport of Dirac fermions in graphene ribbons with zigzag edges. We consider graphene stripes of length $L$ and width $W$. The  potential energy is $U({\bm r})=U_0$ for $0<x<L$. Throughout this section  $\Delta = 10 a_0$, where $a_0=1.42$~\AA~is the carbon-carbon bond length in graphene.

\begin{figure}[!h]
\includegraphics[width=1.0\columnwidth]{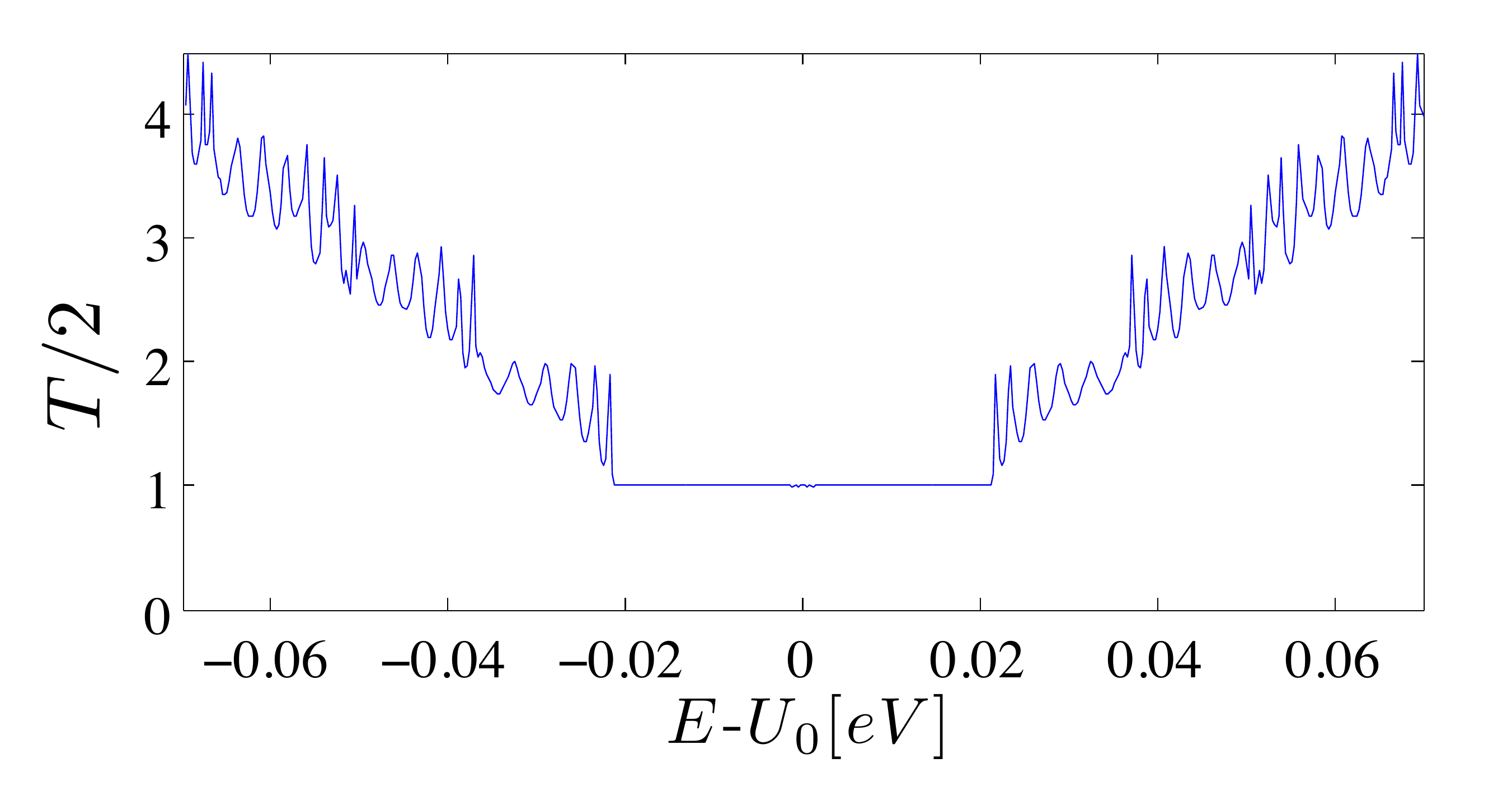}
\caption[]{Transmission coefficient $T=T_K+T_{K^\prime}$ for a graphene ribbon of width $W = 100 \Delta$ and length $L=300\Delta$ as a function of $E-U_0$ in eV.
\label{fig:GvsE-Naive}}
\end{figure}

The conductance is computed using the procedure described in Sec.~\ref{sec:Landauer}. Figure \ref{fig:GvsE-Naive} shows the transmission coefficient $T=T_K+T_{K^\prime}$ as a function of the energy $E$ prior to the spurious mode elimination. We obtain $T_K=T_{K^\prime}$. 

The conductance around the charge neutrality point $E-U_0=0$ is twice the expected value.  Instead of a single open mode, corresponding to edge state in the zigzag boundary, \cite{BreyFertig06} we observe two open modes at low energies. The ``extra" mode is just the spurious one discussed above. 

\begin{figure}[!h]
\begin{center}
\includegraphics[width=1.03\columnwidth]{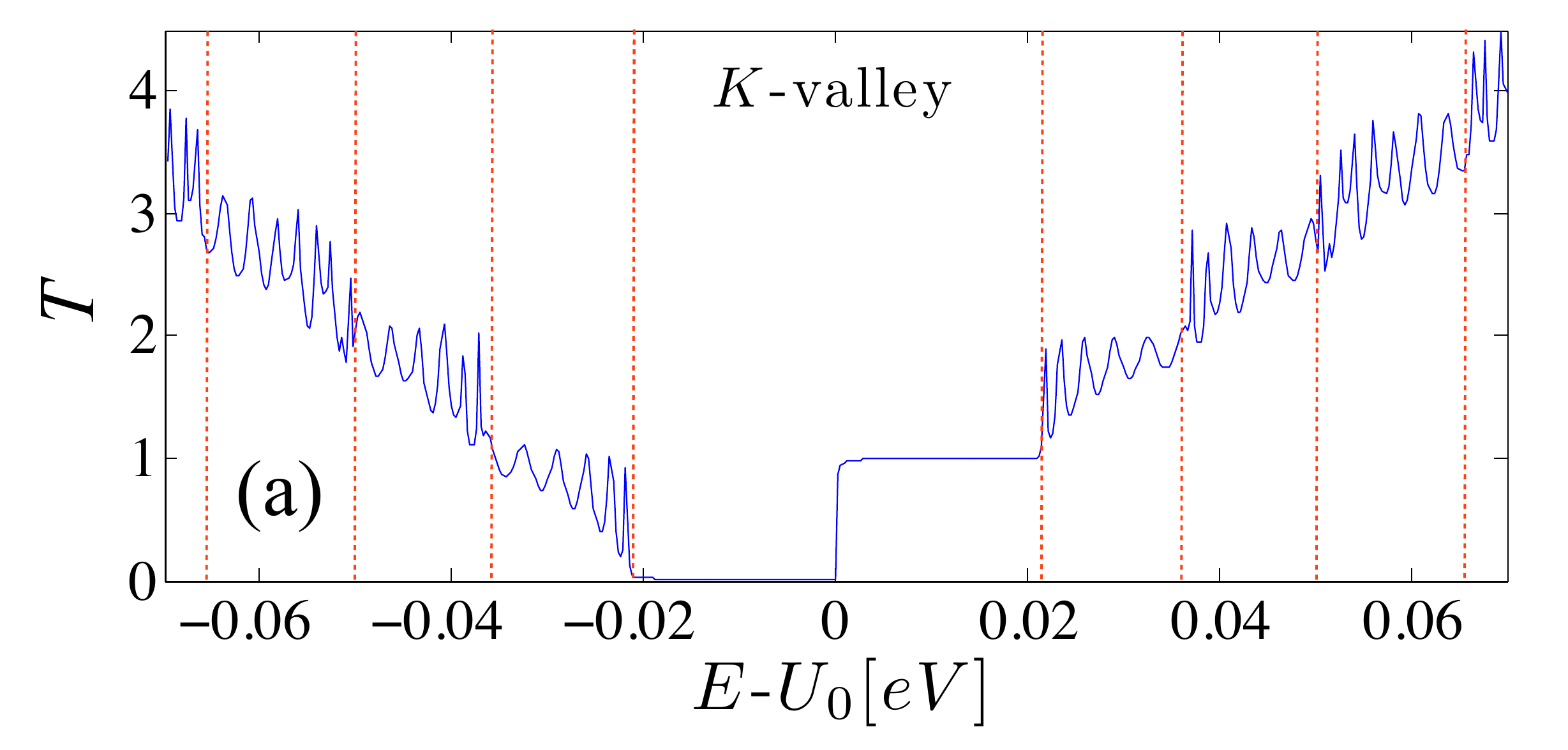}\\
\includegraphics[width=1.03\columnwidth]{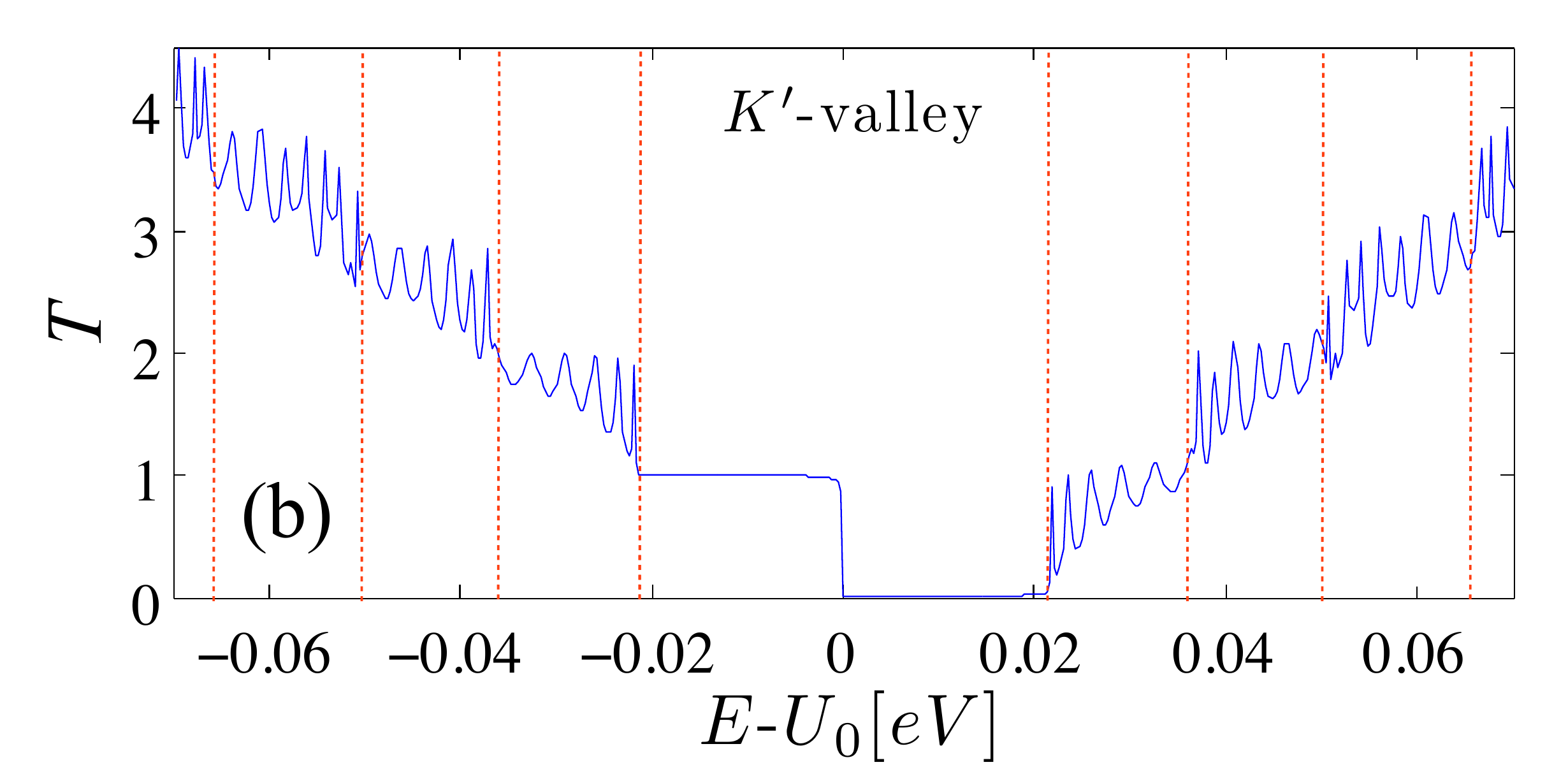}
\end{center}
\vskip-0.5cm
\caption[]{Transmission $T$ for graphene ribbons of width $W = 100 \Delta$ and length $L=300\Delta$ as a function of $E-U_0$ in eV using a massless Dirac equation centered at the (a) $K$-point and (b) $K^\prime$-point. The vertical dashed lines are placed at the threshold energies of the propagating modes in the depicted energy interval.
\label{fig:KKl}}
\end{figure}

The spurious mode is removed following the recipe indicated in Sec.~\ref{sec:spurious}. Figure \ref{fig:KKl} shows that  the transmission coefficients for the $K$ and $K^\prime$-valleys are no longer degenerate. 
At low energies the transfer matrix method gives a perfect unit conductance for $E>0$ around $K$ (the same for $E<0$ around $K^\prime$.) On the other hand, subtracting the corresponding unit step from $T_K$ and $T_{K^\prime}$ both transmission coefficients coincide and show particle-hole symmetry. The total conductance is given by the sum of the contributions from both valleys. 

The low energy features of the conductance can be understood by inspecting the band structure of a zigzag nanoribbon, as shown in Fig.~\ref{fig:DFbandstructure}. For energies $E$ close to the charge neutrality point, after eliminating the spurious mode, there is a single allowed energy band. For $E>0$, right moving states belong to the $K$-valley. Conversely, left moving states belong to the $K^\prime$-valley. The $K^\prime$-valley dispersion relation can be obtained by making $k\rightarrow -k$ in Fig.~\ref{fig:DFbandstructure}. In our geometry, the drain is at the left, while the source reservoir is connected to the right contact. Hence, for $E>0$ in the vicinity of $E=0$ there is a single open channel, which belong to the $K$-valley. The picture is reversed for $E<0$.  


As $|E-U_0|$ is increased, Fig.\ \ref{fig:KKl} shows a sequence of steps decorated by Fabry-Perot-like oscillations. The dashed lines indicate the energy threshold for  successive subband openings inferred from Fig.~\ref{fig:DFbandstructure}. We observe a good agreement between the dashed line energies and the steps in Fig.~\ref{fig:KKl}. The mismatch between the propagation energy in the graphene sheet and the contacts gives raise to the Fabry-Perot-like interferences. This assertion can be quantitatively tested by calculating $T$ for ribbons with a fixed width $W$ for different lengths $L$. We observe that, as $L$ is increased, the Fabry-Perot interference oscillate more rapidly, scaling with $L^{-1}$ (figure not shown). 

Due to their rough edges, such Fabry-Perot interference patterns are unlikely to be observed in lithographic graphene nanoribbons. The situation is more promising in smooth edge graphene ribbons, made by unzipping carbon nanotubes. \cite{Wang2011} 

We note that alternative discretization schemes can be used to construct a transfer matrix to compute the transport of Dirac fermions in ribbons with armchair-like boundary conditions,\cite{BreyFertig06} and possibly also for chiral boundary conditions. \cite{Akhmerov2008} Our motivation to focus only at the zigzag case is twofold: ($i$) In distinction to the others, this boundary condition does not mix the valleys isospins. This allows to solve the problem in the $K$ and $K^\prime$-valleys separately, which speeds up the computation by a factor about $2^3$. ($ii$) The method was put forward to compute transport properties of realistically large samples in the presence of long range disorder, for which the sample boundaries should not play a key role.

\section{Mesoscopic effects}
\label{sec:diffusive}

The electronic transport in graphene at the charge neutrality point has been the subject of intense experimental and theoretical work, as reviewed, for instance in Refs.~\onlinecite{Mucciolo2010,DasSarma2011}. While it is by now established that in good mobility graphene samples the observed  conductivity $ \sigma $ is of the order of $e^2/h$, its value still depends on sample disorder. In the diffusive regime, where the system size $L$  is much larger than the electron mean free path $\ell$, $ \sigma $ is expected to show a weak system size dependence. In the ballistic case, where $L \alt \ell$, the conductivity depends strongly on the system geometry and it is more appropriate to describe transport properties by the conductivity. It is  likely that the graphene samples used in many of the reported experiments correspond to a ballistic-diffusive crossover. \cite{Mucciolo2010} 

In this Section we apply the finite difference method to compute the graphene  conductivity $\sigma $ at the charge neutrality point in the presence of long range disorder. We obtain the expected anti-weak localization relation between $ \sigma$ and the system size, \cite{Hikami92} namely, $\langle \sigma \rangle = (g_sg_v e^2/\pi h) \ln (L/\ell)$. Next, we study the conductance fluctuations at the charge neutrality point and show that they are not universal. 

The transfer matrix approach gives the two-terminal conductance $G$. The conductivity is obtained from the relation $\sigma=(L/W) \langle G \rangle$, where $\langle \cdots \rangle$ stands for a disorder ensemble average. Since we assume that there is no intervalley scattering, there is no need to choose a finite correlation length for the potential fluctuations, \cite{Tworzydlo08} in distinction to other previous theoretical analysis.  \cite{Bardarson2007,Nomura2007,Ryu2007,San-Jose2007,Lewenkopf2008,Mucciolo2010,Rycerz2007,Adam2009} As in Ref.~\onlinecite{Tworzydlo08} we let the potential at each lattice point fluctuate independently. More specifically, we draw $U_{m,n}$ from an uniform distribution within the interval $[-\delta U, \delta U]$.

We compute the conductivity for $10^3$ realizations for different values of $\delta U$, corresponding to  different mean free paths. We assume that the conductivity follows the relation \cite{Tworzydlo08}
\begin{equation}
\label{eq:sigma_corrected}
\frac{\sigma}{G_0} =\frac{1}{\pi} \ln\left[\frac{L}{\ell^*(\delta U)}\right] + C(\delta U) \frac{\Delta}{L},
\end{equation}
where $G_0=g_sg_v e^2/h$.
Here the first term at the right-hand side is just the weak anti-localization relation with $\ell^*(\delta U)$ associated with $\ell$, while the second one accounts for finite size corrections and/or a contact resistance. For every set of disorder realizations with a given value of $\delta U$ we vary the system size ($L= W = 30,60,90, \mbox{and }120\Delta$) and compute $\langle G (L) \rangle$. With the help of \eqref{eq:sigma_corrected}, these data are used to find $\ell^*(\delta U)$ and $C(\delta U)$.

\begin{figure}[!ht]
\begin{center}
\includegraphics[width=0.7\columnwidth]{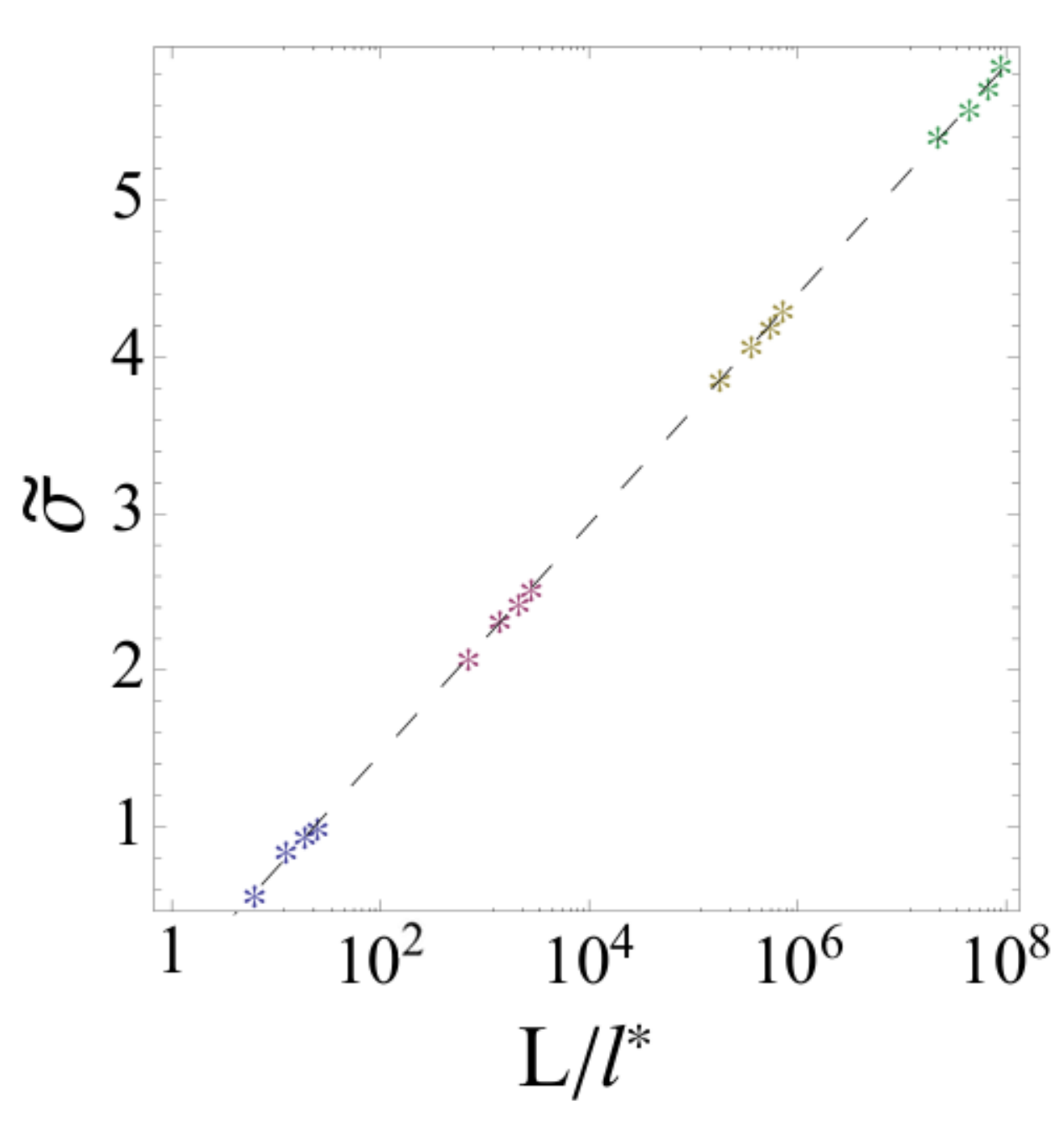}
\end{center}
\vskip-0.3cm
\caption[]{Conductivity $\widetilde{\sigma}$ at the charge neutrality point as a function of system size $L/\ell^*$. Points correspond to different system sizes $L= W = 30,60,90, \mbox{and }120\Delta$ and disorder strengths $\delta U=2,4,6,\mbox{and }8$. The dashed line is given by Eq.~\eqref{eq:sigmatilde}. The lattice spacing is $\Delta =5 a_0$.
\label{fig:AverCond}}
\end{figure}

Figure \ref{fig:AverCond} shows 
\begin{equation}
\widetilde{\sigma} = \frac{1}{\pi} \ln\left[\frac{L}{\ell^*(\delta U)}\right] 
\label{eq:sigmatilde}
\end{equation} 
obtained from the simulations described above. We remark that our method allows to explore a much larger range of $L/\ell^*$ than any previous study.

The current understanding of the conductance fluctuations is less clear. Analytical works make predictions of universal conductance fluctuations (UCF) in graphene. \cite{Kharitonov2008,Kechedzhi2008} However, since these results rely on diagrammatic expansions in powers of $(k_F\ell)^{-1}$, they are applicable for large doping but are hardly relevant to explain the fluctuations at the vicinity of the charge neutrality point. The available experimental data \cite{expUCF} do not provide a very consistent picture, possibly indicating an absence of UCF in graphene at zero doping.   

An early numerically work\cite{Rycerz2007}  studied  conductance fluctuation in graphene in the presence of short range disorder. More recently,  a systematic analysis of the conductance fluctuations in graphene sheets has been performed. \cite{Rossi2011} The study reports  conductance fluctuations as a function of doping, system size $L/\xi$, and disorder strength. The authors consider graphene sheets with transversal periodic boundary conditions (no edges) and use the finite difference method put forward by Tworzydlo and collaborators.\cite{Tworzydlo08} 

We analyze the variance of the conductance fluctuations, var$(G)= \langle G^2\rangle - \langle G\rangle^2$, at the charge neutrality point as a function of $L\ell^*$. In distinction to Ref. \onlinecite{Tworzydlo08}, we find no evidence of universal conductance fluctuations, even going deep into the diffusive regime $L/\ell^*\gg 1$. As stated above, we believe that the strong discrepancy between our results and the prediction from  diagrammatic perturbation theory,\cite{Kharitonov2008,Kechedzhi2008}  is due to the breakdown of the perturbation theory when $k_F \ell \ll 1$, that is, when one approaches the charge neutrality point. In such situation, numerical simulation such as the one presented here give the correct way to approach the problem.

\begin{figure}[!ht]
\begin{center}
\includegraphics[width=0.75\columnwidth]{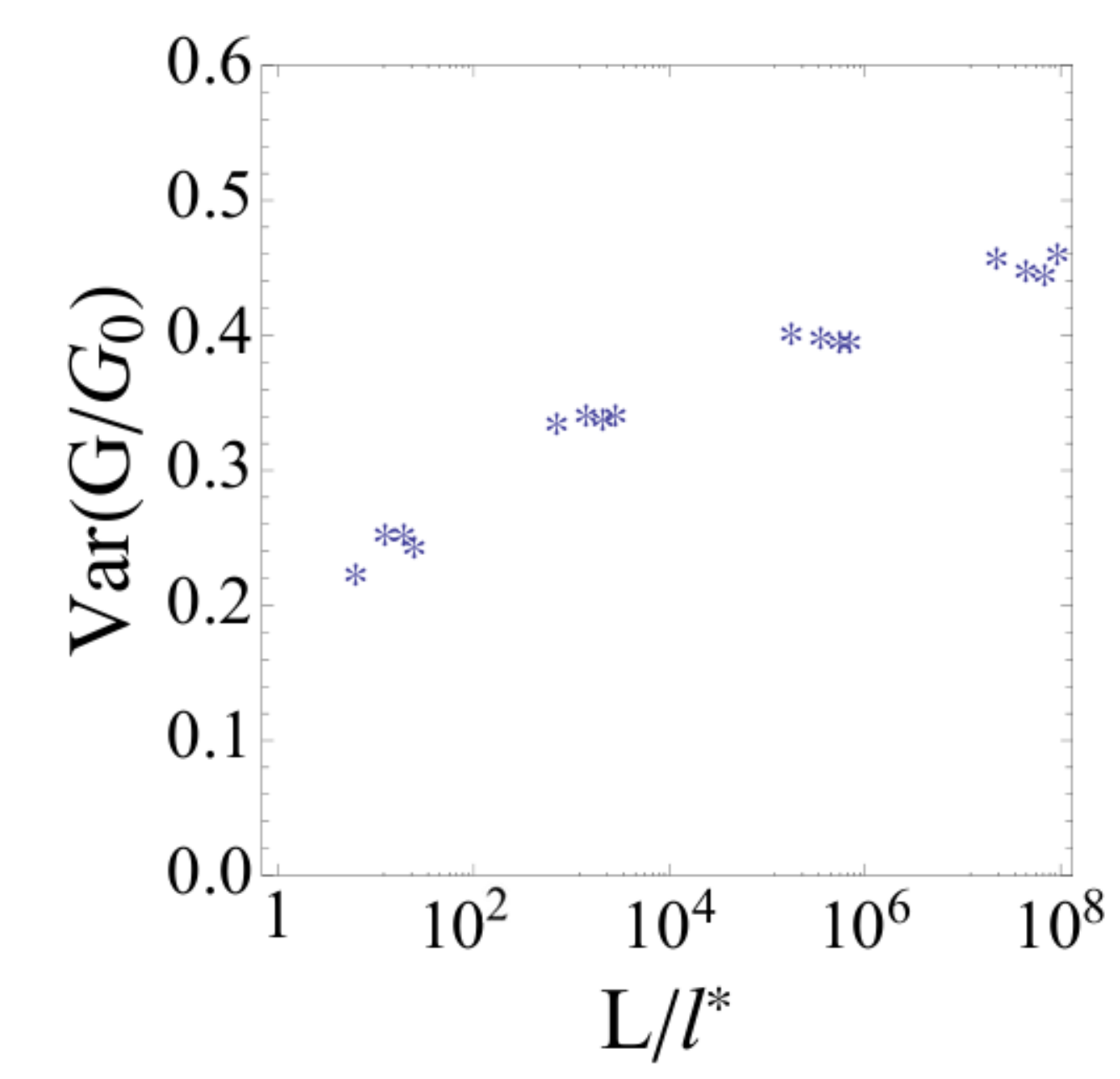}
\end{center}
\caption[]{Conductance fluctuations variance var$(G/G_0)$ as a function of system size $L/\ell^*$.
\label{fig:UCF}}
\end{figure}

\section{Inclusion of a magnetic field}
\label{sec:magnetic}

In this section we show how to modify the transfer matrix discretization scheme presented in Sec.~\ref{sec:discretization} to account for the presence of an external magnetic field. This allows one to calculate (in the single-particle picture) the conductance of ballistic and long range disordered graphene sheets for arbitrary values of a perpendicularly applied magnetic field.

As standard, an external magnetic field is introduced in the Dirac Hamiltonian by minimal coupling. \cite{Goerbig2011} In what follows, we consider the  vector potential ${\bm A}=(-By,0,0)$ that corresponds to the constant magnetic field  ${\bm B} = B \hat{\bm z}$. It is important to mention that, since the discretization schemes along the $x$ and $y$ directions are different, gauge invariance is broken and the choice of ${\bm A}$ has to be cautiously made. The minimal coupling substitution ${\bm p} \rightarrow {\bm p} - (e/c){\bm A}$ modifies Eq.~\eqref{eq:Dirac2}  to
\be
\label{eq:DiracB}
(\partial_x - ieBy/\hbar c) \Psi = -i(\sigma_z\partial_y + \sigma_x V/\Delta)\Psi\;.
\ee
Note that the magnetic field does not couple different valley pseudospin projections. Equation \eqref{eq:DiracB} refers to the $K$-valley. As before, by replacing $\sigma_y \rightarrow - \sigma_y$ one obtains the corresponding equation for the $K^\prime$-valley. 

The transfer matrix discretization for Eq.~\eqref{eq:DiracB} is constructed in the same way as the one presented in Sec~\ref{sec:discretization}. The expression for ${\mathbb M}_m$ is given by Eq.~\eqref{eq:M}, but now ${\mathbb X}_m$ is given by
\be
{\mathbb X}_m={\mathbb J}^{-1} \left( \begin{array}{cc}{\mathbb N}^+-{\mathbb I} - \frac{i\pi  \phi}{2\phi_0}{\mathbb D}\;{\mathbb J}_+ & \widetilde{\mathbb V}_m \\
{\mathbb V}_m & {\mathbb N}^--{\mathbb I} - \frac{i\pi  \phi}{2\phi_0}\bar{{\mathbb D}}\;{\mathbb J}_-\end{array}\right)
\ee
where $\phi=B\Delta^2$ is the magnetic flux through a square unit cell (lattice $A$ or $B$), $\phi_0=h/2ec$ is the magnetic flux quantum, while ${\mathbb D}$ and $\bar{{\mathbb D}}$ are diagonal matrices, whose matrix elements read ${\mathbb D}_{nn^\prime}=n\delta_{nn^\prime}$ and $\bar{{\mathbb D}}_{nn^\prime}=(n-1/2)\delta_{nn^\prime}$.

To illustrate the results that can be obtained by this method, we compute the Landauer conductance of a ballistic graphene sheet  in the anomalous quantum Hall regime. 

\begin{figure}[ht]
\includegraphics[width=0.85\columnwidth]{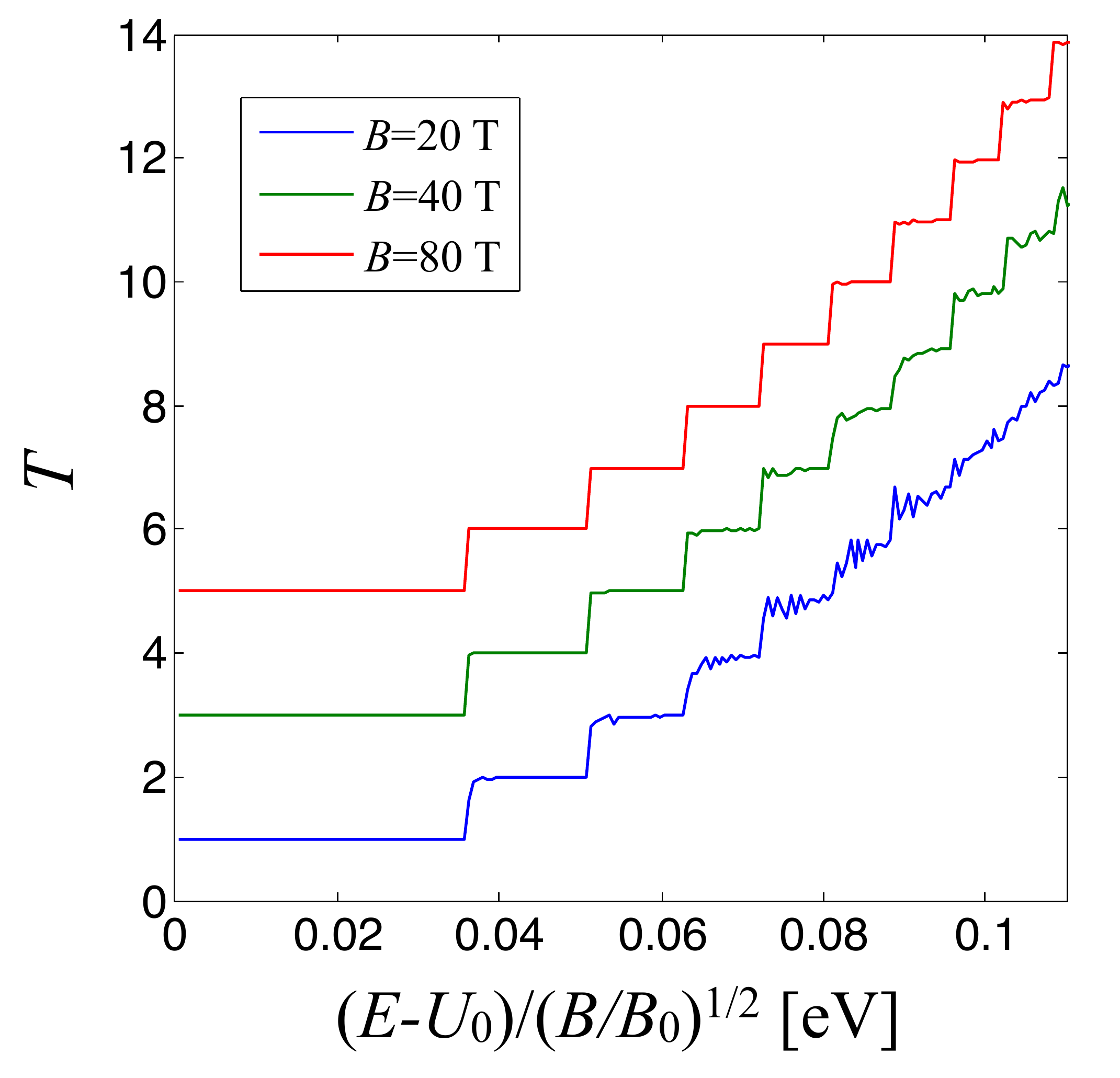}
\caption{Transmission $T$ for different magnetic field values as a function of $E$ for $L= 400a_0$ and $W=200a_0$. To make the magnetic field dependence of the Landau levels in graphene explicit, the energy axis is scaled by $(B_0/B)^{1/2}$, with $B_0=20 T$. After scaling all curves largely coincide. For the sake of clarity, the transmission for $B=40$ T is offset by +2 units, while $T$ for $B=80$ T is offset by +4 units. }
\label{fig:QHE}
\end{figure}

Figure \ref{fig:QHE} shows the $K$-valley conductance for a rectangular graphene ribbon geometry, where $L =400\Delta$, $W= 200\Delta$ and $\Delta = a_0$, subjected to different perpendicular $B$ fields. The energy axis is scaled by $(B_0/B)^{1/2}$, where $B_0=20$ T to make evident that  the Landau level energies $E_N$ scale with $B^{-1/2}$, as expected for Dirac fermions. \cite{CastroNeto09} More precisely, $E_N=\hbar \omega_c N^{1/2}$, where the cyclotron frequency $\omega_c = v_F(2eB/\hbar)^{1/2}$. The transmission steps in Fig.~\ref{fig:QHE} nicely coincide with the Landau level energies. The different $B $-field conductance curves strongly overlap. We observe that as $E$ is increased, the conductance steps become less well-defined. This behavior is particularly evident for $B=20$ T and become less pronounced as $B$ is increased. The interpretation is that with increasing $B$, the edge states become more localized and hence tunneling between edges, which destroys quantization, is suppressed. 

 For the system we analyze, the $K^\prime$-valley transmission coefficient for $E-U_0>0$ is $T_{K^\prime} = T_K -1$. The corresponding figure is not shown, but bears a strong analogy to the situation we discussed in Sec.~\ref{sec:ballistic}.  
The full conductance is given by the contribution of both $K$ and $K^\prime$ valley pseudospin components, namely $G= (g_s e^2/h) (T_K + T_{K^\prime})$ and consistent with the result found in the literature. \cite{CastroNeto09,Goerbig2011}

%

\section{Conclusion and outlook}
\label{sec:conclusion}

In this paper we put forward a numerical method to calculate transport properties of two-dimensional massless Dirac particles. The numerical procedure   combines a finite difference description of the Dirac equation with a current conserving transfer matrix formalism. The method is useful to study the Landauer conductance in graphene and surface states in 3D topological insulators.

The use of a finite difference description of the Dirac equation to calculate transport coefficients under given boundary conditions demands tailormade discretization schemes. We consider a system with a rectangular geometry. In the longitudinal direction, parallel to the axis connecting source and drain, a Stacey-like discretization allows us to implement the transfer matrix formalism. In the transversal direction, a Susskind-like discretization makes possible the inclusion of zigzag-like edge boundary conditions. By construction, our method avoids the fermion doubling problem.


The proposed method becomes numerically advantageous with respect to the standard ones, that employ an atomic basis, when treating systems where long range disorder is dominant. In this case, the disorder range $\xi$ sets the discretization length scale. That is, $\Delta \alt \xi$ instead of $\Delta \sim a_0$ as in any atomistic basis discretization. For a ribbon of length $L$ and width $W$, the computation time of transport properties scales with $(L/\Delta) \times(W/\Delta)^3$. Hence, for a given geometry, our discretization method is of the order of $(\xi/a_0)^4$ times computationally faster than standard atomistic ones.

To illustrate how the method works, we analyzed the transport properties of ballistic and diffusive graphene ribbons in the absence of magnetic field, as well as the anomalous quantum Hall plateaus in a graphene sample under a strong perpendicular magnetic field.

We believe that the most promising features of our method is that it allows considering different kinds of boundary conditions. The present study addresses the case of zigzag-like edges, but the transversal discretization can be modified to mimic arm-chair (see Appendix \ref{app:armchair}) and chiral edges, as well as non-trivial periodic boundary conditions suitable to describe  surface states of certain topological insulators. We believe our method can be very helpful for the study of issues related to the intriguing nature of boundaries in Dirac systems.

\begin{acknowledgments}
We thank Eduardo Mucciolo and Nancy Sandler for useful discussions. This work is supported in part by the Brazilian funding agencies CNPq and FAPERJ.
\end{acknowledgments}

\appendix
\section{Armchair graphene ribbons}
\label{app:armchair}

In this Appendix we modify the discretization procedure presented in Sec.~\ref{sec:discretization} to address graphene 
ribbons with armchair-like edges. To satisfy the boundary conditions of such systems one needs to mix both $K$ and $K^\prime$ Dirac-valley components, \cite{BreyFertig06,CastroNeto09} namely, 
\be
\Psi= \left(\begin{array}{c}\psi\\ \widetilde{\psi} \end{array}\right) \quad \mbox{and} \quad
\Psi^\prime=  \left(\begin{array}{c}\psi^\prime\\ \widetilde{\psi}^\prime \end{array}\right).
\ee
The armchair-like boundary conditions at the graphene ribbon edges are accounted for by taking \cite{BreyFertig06} 
\be 
\label{eq:edge1}
\Psi ({\bm r})\Big|_{x=0}=\Psi^{\prime} ({\bm r})\Big|_{x=0} 
\ee 
and  
\be
\label{eq:edge2}
\Psi ({\bm r})\Big|_{x=W}=e^{i\delta\! K W}\Psi^{\prime}({\bm r})\Big|_{x=W} ,
\ee
where $\delta K=4\pi/(3a_0)$ and $a_0$ is the graphene lattice constant. Here we chose the $x$-axis as the transversal direction and the $y$-axis as the electron propagation direction. Equations \eqref{eq:edge1} and \eqref{eq:edge2}  guarantee that the wave function amplitude vanishes on both sublattices at the ribbon edges $x=0$ and $x=W$.\cite{BreyFertig06}

In distinction to the zigzag case, described in terms of a single-valley Dirac equation given by Eq.~\eqref{eq:H_Dirac}, here we have to solve the effective 4-component spinor Dirac Hamiltonian 
\begin{align}
-i\hbar v \left( \!\begin{array}{cc} 
\sigma_x \partial_x + \sigma_y \partial_y &0\\
0&\sigma_x \partial_x - \sigma_y \partial_y  \end{array}\! \right)
\left(\!\begin{array}{l}\Psi\\ \Psi^\prime \end{array}\!\right)
=
E \left(\!\begin{array}{l}\Psi\\ \Psi^\prime \end{array}\!\right),
\end{align}
with boundary conditions given by Eqs.~\eqref{eq:edge1} and \eqref{eq:edge2}. 

The problem is cast in the discrete space as follows: The fields $\Psi$ and $\Psi^\prime$ are evaluated at 
\begin{align}
\label{eq:lattice_ac}
&\psi_m(x= n \Delta)  \hskip0.5cm  n=0,\cdots, N-1 \nonumber \\
&\tilde{\psi}_m (x= n \Delta)\hskip0.5cm    n=1,\cdots, N   \nonumber \\
&\psi^{\prime}_m (x= n \Delta)\hskip0.5cm  n=1,\cdots ,N &\nonumber \\
&\tilde{\psi}^{\prime}_m (x= n \Delta)\hskip0.5cm     n=0,\cdots, N-1\;;    
\end{align}
where $\Delta$ is the lattice spacing and $m$ labels the mesh position, $y = m \Delta$. The armchair-like edge boundary conditions \eqref{eq:edge1} and \eqref{eq:edge2} read
\begin{align}
\label{eq:bc_ac}
\psi_m (x=N\Delta&) = e^{i\delta \! K W} \psi_m^{\prime} (x= N \Delta), \nonumber \\
\tilde{\psi}_m (x= 0&)  =  \tilde{\psi}_m^{\prime} (x= 0 ),  \nonumber \\
\psi_m^{\prime} (x= 0&) = \psi_m (x= 0 ),  \nonumber \\
\tilde{\psi}_m^{\prime} (x= N \Delta&)  =  e^{-i\delta\! K W} \tilde{\psi}_m (x= N \Delta),
\end{align}
where $W=(N+1/2)\Delta$, as in Ref. \onlinecite{BreyFertig06}. We set $\Delta=a_0$ to make easy the comparison with the tight-binding model results.\cite{comDelta} 

Following the steps of the discretization procedure described in Sec.\ \ref{sec:discretization}, Eqs. \eqref{eq:lattice_ac} and \eqref{eq:bc_ac} lead to a modified ${\mathbb X}$-matrix given by
\be \small
\label{eq:X-AC}
{\mathbb X}_m={\mathbb J}^{-1} \left( \begin{array}{cccc}-({\mathbb I} - {\mathbb N}^+) & \widetilde{\mathbb V}^m &0&{\mathbb A} \\
{\mathbb V}^m & -({\mathbb I} - {\mathbb N}^-) &{\mathbb B}&0 \\ 
0&{\mathbb A}^{\dagger}& -({\mathbb I} - {\mathbb N}^+) & \widetilde{\mathbb V}^m \\ 
{\mathbb B}^{\dagger}&0& {\mathbb V}^m & -({\mathbb I} - {\mathbb N}^-) \end{array}\right)
\ee
where ${\mathbb A}$ and ${\mathbb B}$ are matrices whose single non-zero elements are ${\mathbb A}_{N,N}=-e^{-i\delta K W}$ and ${\mathbb B}_{1,1}=-1$, respectively. Similarly, the ${\mathbb J}$-matrix reads
\be
\label{eq:J-AC}
{\mathbb J}= \left( \begin{array}{cccc} {\mathbb I} + {\mathbb N}^+ & 0 &0&{\mathbb A} \\
0 & {\mathbb I} + {\mathbb N}^- &{\mathbb B}&0 \\ 
0&{\mathbb A}^{\dagger}& {\mathbb I} + {\mathbb N}^+ & 0 \\ 
{\mathbb B}^{\dagger}&0& 0 & {\mathbb I} + {\mathbb N}^- \end{array}\right).
\ee
The transfer matrix ${\mathbb M}$ is defined, as standard, by
\begin{align}
\left(\begin{array}{c} {\bm \Psi}_{m +1} \\ {\bm \Psi}^\prime_{m+1} \end{array}\right)
=  {\mathbb M}_m
\left(\begin{array}{c} {\bm \Psi}_{m } \\ {\bm \Psi}^\prime_{m} \end{array}\right).
\end{align}
As in Sec.~\ref{sec:discretization}, ${\mathbb M}_m$ is related to ${\mathbb X}_m$ by Eq.~\eqref{eq:M}. 

Due to translational invariance, the transfer matrix of pristine armchair graphene ribbons does not depend on the index $m$, that is ${\mathbb M}_{\rm PUC}={\mathbb M}_m$.  For a given energy $E$, ${\mathbb M}_{\rm PUC}$ has a subset of eigenvalues of the form $e^{i k_a \Delta}$, with $k_a$ real. Those correspond to propagating modes and allow one to infer the system band structure. For a detailed discussion see Sec.~\ref{sec:spurious}.

\begin{figure}[h!]
\vskip0.1cm
\begin{center}
\includegraphics[width=\columnwidth]{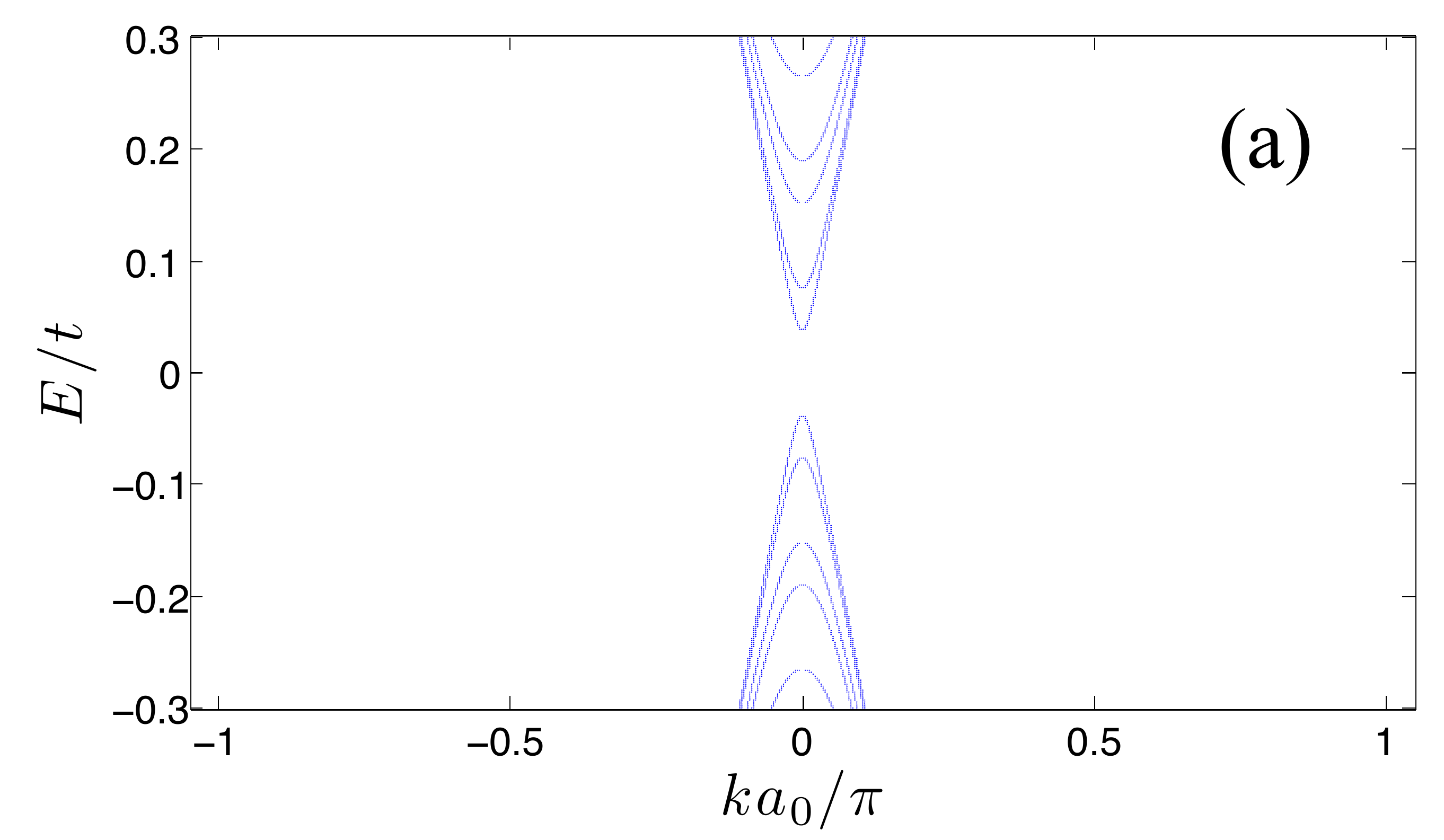}
\includegraphics[width=\columnwidth]{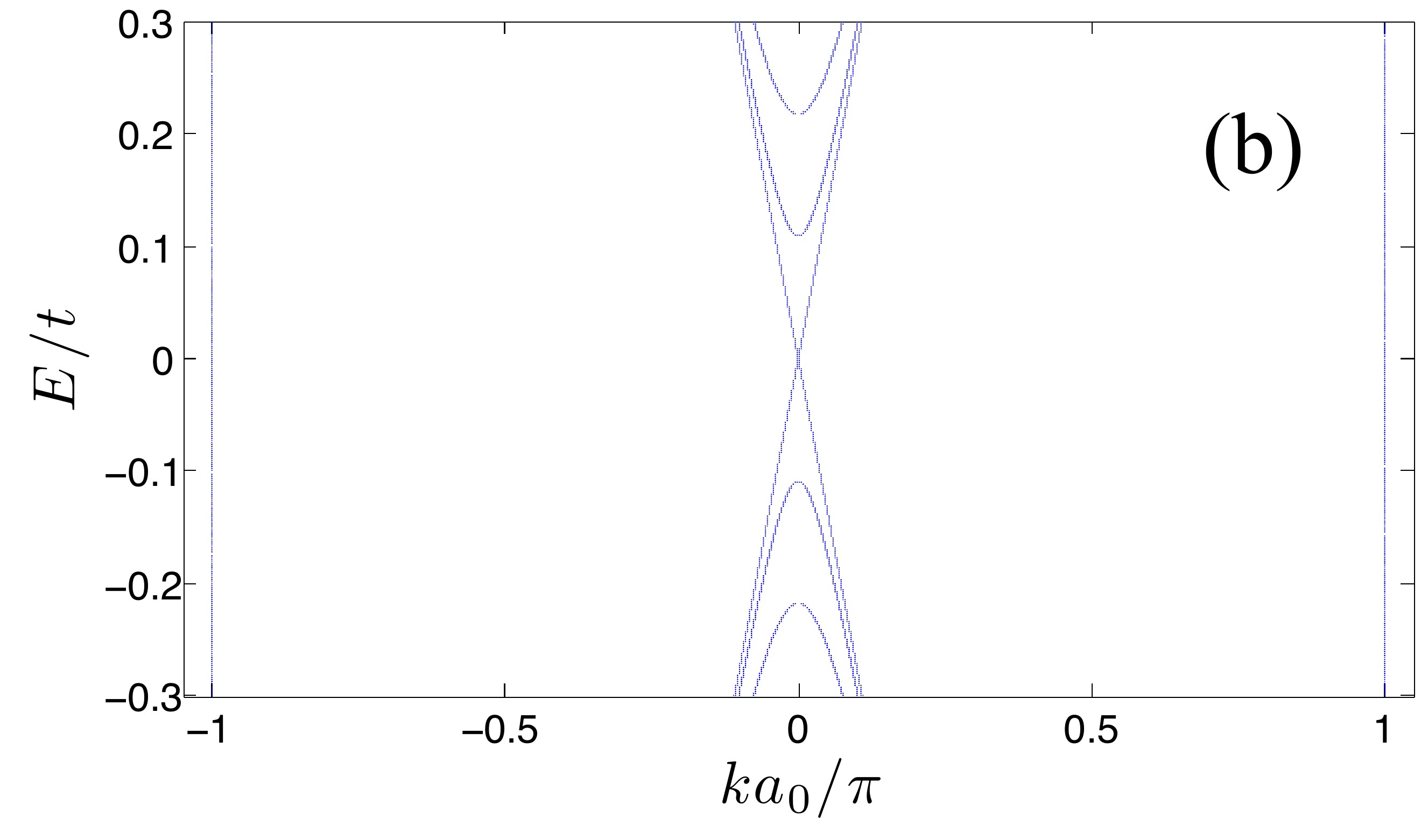}
\end{center}
\vskip-0.3cm
\caption{Dispersion relation $E/t$ as a function of $ka_0/\pi$ for graphene ribbons with armchair-like edges, where $t$ is the graphene tight-binding nearest-neighbor hopping integral. (a) Insulator ribbon, $N=24$. (b) Metallic ribbon, $N=25$. }
\label{fig:ACband}
\end{figure}

The described discretization scheme reproduces with good acuracy the low-energy nearest neighbor tight-binding dispersion relation for graphene ribbons with armchair edges. As $N$ is increased, we obtain the correct sequence of two insulator band structures followed by a single metallic one. Figures~\ref{fig:ACband}a and \ref{fig:ACband}b show the band structure for typical metallic and insulator ribbons, respectively. As in the zigzag case, there are spurious modes in the metallic ribbon, whose wave functions oscillate on the scale of the lattice spacing. Those must be eliminated to understand the system transport properties. This can be done in a similar way as that discussed in Sec.~\ref{sec:spurious}.

\begin{figure}[h]
\begin{center}
\vskip0.5cm
\includegraphics[width=\columnwidth]{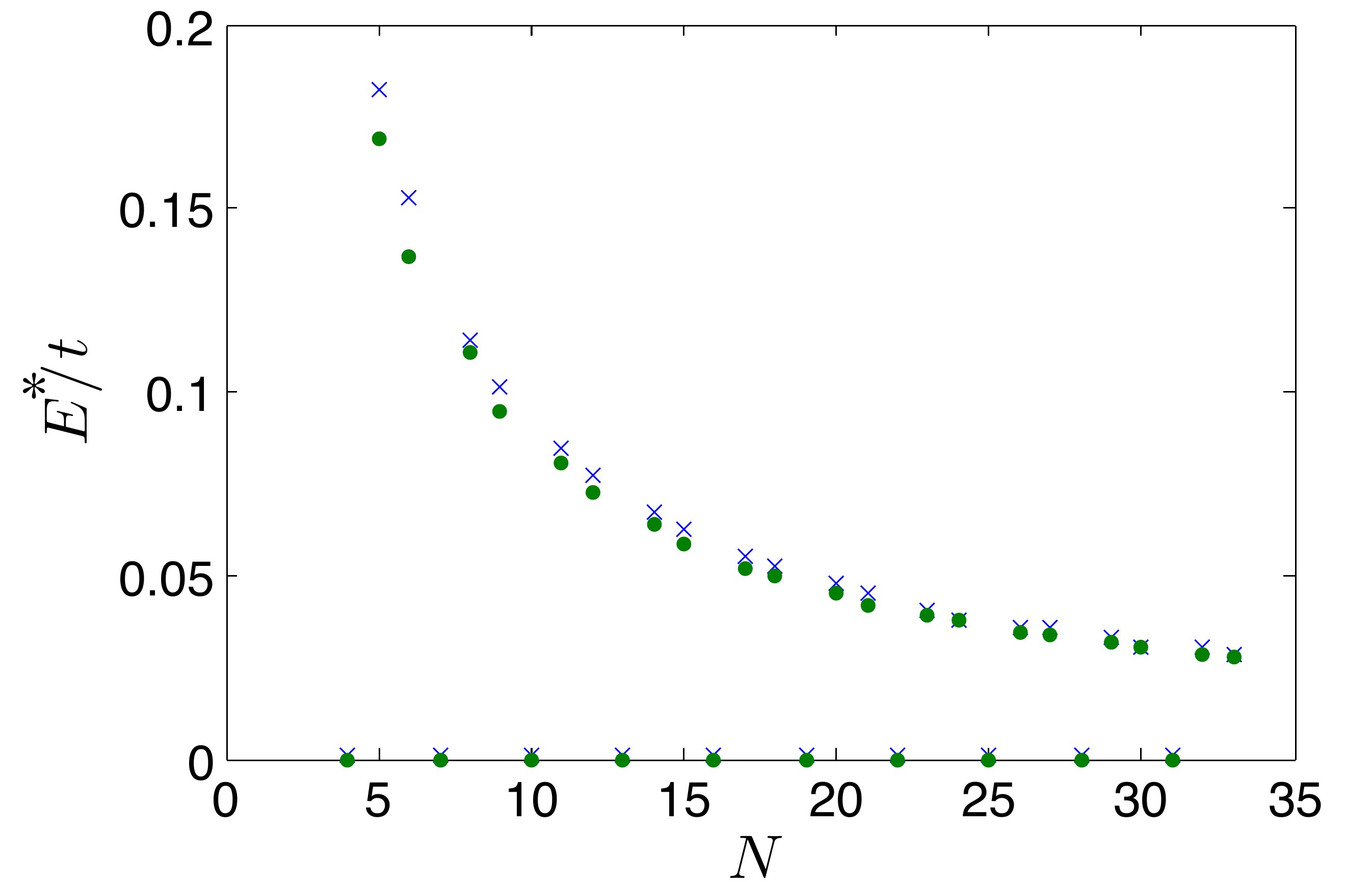}
\end{center}
\caption{Lowest energy $E^*/t$ of the graphene ribbons with armchair-like edges as a function of the ribbon width $W=N a_0$. Crosses correspond to the energies values calculated using the finite difference method, while the dots stand for nearest-neighbor tight-binding model results.}
\label{fig:Gap}
\end{figure}

Figure \ref{fig:Gap} shows the dependence of the lowest confined state energy versus ribbon width given in terms of $N$. For comparison, the corresponding energy obtained from the nearest-neighbor tight-binding model as also shown. The agreement is very good, comparable to the one obtained solving the Dirac equation without discretization. \cite{BreyFertig06}

These results show that, by using the full 4-component graphene effective Dirac Hamiltonian, the transfer matrix method can accurately describe the band structure of graphene ribbons with armchair edges. After the elimination of the spurious modes that appear in the metallic case (not done here), the computation of the conductance follows the procedure described in Sec.~\ref{sec:discretization}. 

Further extensions of the method to deal with other kinds of edge boundary conditions, such as the chiral ones, is also possible. The later requires a modification of both the primitive unit cell and the boundary equations, namely, \eqref{eq:lattice_ac} and \eqref{eq:bc_ac}. 

\section{Charge conservation.}
\label{ap:chargcons}
In order to conserve the total current at any arbitrary ribbon cross section, the discretized current operator must
fulfill the following condition
\be
\langle {\bm \Psi}_{m+1}|J_x| {\bm \Psi}_{m+1} \rangle =
\langle {\bm \Psi}_{m}|J_x| {\bm \Psi}_{m} \rangle.
\ee
As a consequence, one writes
\be
 {\mathbb M}_m^{\dagger}{\mathbb J_x}{\mathbb M_m}={\mathbb J_x},
\ee
which is equivalent to 
\be
{\mathbb J}_x^{-1}{\mathbb M}_m^{\dagger}{\mathbb J}_x={\mathbb M}_m^{-1}.
\ee
Using  ${\mathbb M_m}$ as given by Eq. (\ref{eq:M}), we show that in order to preserve current, ${\mathbb J}_x$ must satisfy
\be
{\mathbb J}_x^{-1}{\mathbb X}_m^{\dagger}{\mathbb J}_x={\mathbb X}_m.
\ee
Indeed, this can be readily shown with the help of Eqs. (\ref{eq:X}), (\ref{eq:J}), and (\ref{eq:Jx}).



\begin{thebibliography}{10}
\bibitem{TI}
The same finite difference scheme can be used for topological insulators. In this case, $\Psi$ evaluated at the lattice points of $A$ and $B$ is a discrete representation of the two spin projections of the electron wave function.
\bibitem{comDelta}
For a given value of the ribbon width $W$, the computational cost can be reduced by decreasing the number of transversal sites $N$ to $N^\prime$, provided $N$ and $N^\prime$ have the same remainder left when divided by 3, that is, $N \equiv N^\prime \,(\mbox{mod} \,3)$. Consequently, the modified lattice parameter is increased to $\Delta^\prime=a_0(N+1/2)/(N^{\prime}+1/2)$.
\end{thebibliography}
\end{document}